\documentclass[11pt]{article}
\usepackage{geometry}                
\usepackage{graphicx}
\usepackage{multirow}
\usepackage{fancyhdr}
\usepackage{lineno}
\usepackage{amsmath}
\usepackage[plain]{algorithm2e}
\usepackage{color}
\usepackage{colortbl}
\usepackage{amssymb}
\usepackage{amsthm}
\newtheorem{proposition}{Proposition}
\usepackage[all,cmtip]{xy}
\usepackage{epstopdf}
\usepackage{authblk}

\title{Inference on tissue transplantation experiments}
\author[1]{Yue Wang\thanks{Corresponding author. E-mail address: yuewang@ihes.fr (Y. Wang).}}
\author[2]{Boyu Zhang}
\author[1]{J\'er\'emie Kropp}
\author[1,3,4]{Nadya Morozova}
\affil[1]{Institut des Hautes \'Etudes Scientifiques (IH\'ES), 91440, Bures-sur-Yvette, France}
\affil[2]{Department of Mathematics, Princeton University, 08544, Princeton, NJ, United States}
\affil[3]{Institute for Integrative Biology of the Cell (I2BC), CEA, CNRS, Universit\'e Paris-Sud, Universit\'e Paris-Saclay, 91198, Gif-sur-Yvette, France}
\affil[4]{Komarov Botanical Institute, Russian Academy of Sciences (BIN RAS), 197376, Saint Petersburg, Russia}
\date{}
\begin{document}
\maketitle

\begin{abstract}
We review studies on tissue transplantation experiments for various species: one piece of the donor tissue is excised and transplanted into a slit in the host tissue, then observe the behavior of this grafted tissue. Although we have known the results of some transplantation experiments, there are many more possible experiments with unknown results. We develop a penalty function-based method that uses the known experimental results to infer the unknown experimental results. Similar experiments without similar results get penalized and correspond to smaller probability. This method can provide the most probable results of a group of experiments or the probability of a specific result for each experiment. This method is also generalized to other situations. Besides, we solve a problem: how to design experiments so that such a method can be applied most efficiently.
\end{abstract}

\smallskip
\noindent \textbf{Keywords.} 

\noindent Experimental result inference; Similarity between experiments; Penalty function; Experimental design.

\smallskip
\noindent \textbf{Highlights.} 

\noindent (1) Review of tissue transplantation experiments for various species.

\noindent (2) A penalty function-based method that infers unknown results of binary experiments.

\noindent (3) Generalized methods for experiments with stochastic results or multiple results.

\noindent (4) Experimental design that maximizes the inference efficiency.

\section{Introduction}
In this paper, we concern tissue transplantation experiments of various species. During the development of embryos, one piece of the donor tissue is excised and transplanted into a slit in the host tissue; then, one observes how the grafted tissue behaves. Since the grafted tissue is placed in an unfamiliar environment, it might be assimilated by the host \cite{arresta2005lens} or even transdifferentiate into a neither-donor-nor-host tissue \cite{henry1987inductive}. The development might also exhibit abnormalities \cite{hamburger1939development}. According to the tissue type and normality, the fate of the grafted tissue can be roughly classified into eight possibilities: develop normally as host tissue; develop abnormally as host tissue; develop normally as donor tissue; develop abnormally as host tissue; develop normally as a third tissue; develop abnormally as a third tissue; develop totally abnormally that cannot determine tissue type; death. For example, if one piece of the \emph{Xenopus laevis} upper lateral lip (developmental stage 11) is transplanted to the lower lip (developmental stage 11), it will develop normally as the lower lip, the host tissue \cite{krneta2010temporal}. The transplanted tissue might induce a new structure (head/base/limb) \cite{newman1974interaction} or even induce a new structure in another species \cite{muneoka1987intercalation,saunders1976inductive}. In some experiments, the results are deterministic, while others are stochastic (e.g., develops normally with probability 60\%).

In developmental biology, a central question is why a zygote (in the correct environment) could develop into a normal adult animal. To understand why the developmental process in a normal environment works, we also need to understand why the developmental process in an abnormal environment does not work. Tissue transplantation experiments describe how tissues behave in abnormal environments. Thus they could provide crucial knowledge of developmental biology.

Just for \emph{Xenopus laevis}, there are around 1000 tissues across around 70 stages \cite{segerdell2008ontology}. Thus there are millions or even billions of possible tissue transplantation experiments, among which only a few have been executed. If we also consider other commonly studied species, there could be trillions of experiments with unknown results. To extend our understanding of tissue transplantation, we need some methods to infer the unknown experimental results based on the known experimental results.

Besides the known experimental results, there is some common sense in biology that might help infer unknown results. Some tissue pairs are more similar than others. We expect that transplantations between similar tissues are more likely to produce normal results. Similar experiments (similar hosts and similar donors) tend to have similar results.

With such knowledge, the experiments can be represented by a graph, where each experiment is a node, and similar experiments are linked by edges. Each node has a label, namely the experimental result. Now the problem is to infer partially observed labels on a graph. 

We adopt a penalty function that evaluates the guesses of experimental results according to the graph structure. Using this method, we can obtain the most probable results of a group of experiments or the probability of a specific result for each experiment. 

The above method works when the known experimental results are deterministic. For experiments with stochastic results, we can decompose them into deterministic results with different probabilities, apply the above method, and then take the average. Besides, the penalty function can be modified to accommodate experiments with more than two possible results.

These methods conduct inference with given experimental results. A new question is if we can choose some experiments to conduct, and use them to infer other experiments, what is the most efficient choice? We need to guarantee the inference quality and also try to reduce the number of conducted experiments. This problem becomes an experimental design problem, depending on the properties of our inference methods and the similarities between experiments.

In Section \ref{S2}, we review studies on tissue transplantation experiments. In Section \ref{S3}, we develop a method to infer the unknown experimental results, where known experimental results are deterministic and binary. In Section \ref{S4}, we generalize this method to experiments with stochastic results. In Section \ref{S5}, we generalize this method to experiments with more than two results. In Section \ref{Snew}, we develop an experimental design method so that the inference methods can be applied efficiently. We finish with some discussions in Section \ref{S6}.

\section{Summary of transplantation experiments}
\label{S2}
There are many works that concern tissue transplantation experiments on various species: \emph{Xenopus laevis} and \emph{Xenopus Borealis} \cite{arresta2005lens,beazley1975development,beazley1975factors,cameron1977evidence,cannata2008optic,cooke1987organization,dickinson2006development,elliott2013transplantation,gaze1979orientation,gont1993tail,grainger1988reinvestigation,henry1987inductive,hunt1975neuronal,hunt1972development,jacobson1968development,jones1987development,jones1987development2,krneta2010temporal,lee2013early,sater1990role,smith1983dorsalization,tschumi1957growth,tucker1995tail}, chick \cite{hamburger1939development,javois1986handedness,saunders1976inductive,waddington1933induction}, \emph{Hydra attenuata} \cite{newman1974interaction}, \emph{Cancer gracilis} and other crabs \cite{kao1996homeotic,kao1997limb}. We consider a standard paradigm: one tissue that appears in normal development (donor tissue) is transplanted to another tissue (possibly with the removal of some tissues) that appears in normal development (host tissue), then observe what the grafted tissue will develop into (host tissue, donor tissue, or neither), and whether the development is normal or abnormal. We introduce eight major groups of experiments that fit this paradigm.

(1) In the experiments reported by Krneta-Stankic et al. \cite{krneta2010temporal}, \emph{Xenopus laevis} mesoderm and lip tissues were transplanted between each other. The grafted tissues all developed like host tissues, while some of them were normal (denoted by NH), and some were abnormal (denoted by AH). The results are presented in Table \ref{KS}. For example, the entry ``AH'' with host ``AM19'' and donor ``PM15'' means that if we take a piece of presomitic mesoderm at stage 15 (denoted by PM15), and transplant it to anterior paraxial mesoderm at stage 19 (denoted by AM19), the grafted tissue will develop abnormally as the host tissue (AH).

\begin{table}[]
	\begin{tabular}{lllllllll}
		&    &    &    &    & \multicolumn{4}{l}{Donor tissue} \\
		&    & AM19 & PM19 & PM15 & UL11   & LL11   & LL15   & LL19   \\
		& AM19 & ? & NH & AH & AH   & AH   & AH   & NH   \\
		& PM19 & ?  & NH & ?  & NH   & NH   & ?    & ?    \\
		 & PM15 & ?  & ?  & ? & ?    & ?    & ?    & ?    \\
		Host &UL11 & ?  & NH & ?  & NH   & NH   & ?    & ?    \\
		tissue & LL11 & ?  & NH & ?  & NH   & NH   & ?    & ?    \\
		 & LL15 & ?  & ?  & ?  & ?    & ?    & ?   & ?    \\
		& LL19 & ?  & ?  & ?  & ?    & ?    & ?    & ?  \\
		&&&&&&&&
	\end{tabular}
	\caption{\emph{Xenopus laevis} transplantation results reported by Krneta-Stankic et al. \cite{krneta2010temporal}. AM is anterior paraxial mesoderm; PM is presomitic mesoderm; UL is upper lateral lip; LL is lower lip. Number is developmental stage. NH means normal host; AH means abnormal host. Question marks are experiments with unknown results.}
	\label{KS}
\end{table}

(2) In the experiments reported by Henry and Grainger \cite{henry1987inductive}, the presumptive lens ectoderm (which would develop into the lens) of \emph{Xenopus laevis} was removed from the lens-forming region, and different ectoderm tissues were transplanted to this location to check whether the grafted tissue could develop into the lens. The experimental results are stochastic: some cases had lens formation, some did not. The results are shown in Table \ref{HG}. For example, the entry ``61\%'' with host ``LFR$\backslash$PLE14'' and donor ``PLE11'' means that if we transplant a piece of presumptive lens ectoderm at stage 11 (denoted by PLE11) to the lens forming region without presumptive lens ectoderm at stage 14 (denoted by LFR$\backslash$PLE14), there will be 61\% cases with lens formation. 

\begin{table}[]
	\begin{tabular}{lllllll}
		&                          &       &       & \multicolumn{3}{l}{Donor tissue} \\
		&                          & PLE11 & PLE12 & PLE14     & PLE16     & PLE19    \\
		& LFR\textbackslash{}PLE14 & \,  61\%  & \,  58\%  & \,  82\%      &  \,  \, ?         &  \,  \, ?        \\
		\begin{tabular}[c]{@{}l@{}}Host\\ tissue\end{tabular} & LFR\textbackslash{}PLE16 &  \,  \, ?     &  \,  \, ?     &  \,  \, ?         &  \,  \, ?         &  \,  \, ?        \\
		& LFR\textbackslash{}PLE19 & \,  \, 4\%   & \,  24\%  & \,  83\%      &  \,  \, ?         & 100\%    \\
		&                          &       &       &           &           &          \\
		&                          &       &       & \multicolumn{3}{l}{Donor tissue} \\
		&                          & AVE11 & AVE12 & AVE14     & AVE16     & AVE19    \\
		& LFR\textbackslash{}PLE14 & \,  29\%  & \,  50\%  & \,  14\%      & \,  \,  0\%       & \,  \,  0\%      \\
		\begin{tabular}[c]{@{}l@{}}Host\\ tissue\end{tabular} & LFR\textbackslash{}PLE16 &  \,  \, ?     &  \,  \, ?     &  \,  \, ?         &  \,  \, ?         &  \,  \, ?        \\
		& LFR\textbackslash{}PLE19 & \,  \,  8\%   & \,  13\%  & \,  \,  0\%       &  \,  \, ?         & \,  \,  0\%      \\
		&                          &       &       &           &           &          \\
		&                          &       &       & \multicolumn{3}{l}{Donor tissue} \\
		&                          & PVE11 & PVE12 & PVE14     & PVE16     & PVE19    \\
		\begin{tabular}[c]{@{}l@{}}Host\\ tissue\end{tabular} & LFR\textbackslash{}PLE14 &  \, 11\%  & \,  16\%  &  \,  \, 4\%       &  \,  \, 0\%       &  \,  \, ?      \\
		&&&&&& 
	\end{tabular}
	\caption{\emph{Xenopus laevis} transplantation results reported by Henry and Grainger \cite{henry1987inductive}. PLE is presumptive lens ectoderm; AVE is anterior ventral ectoderm; PVE is posterior ventral ectoderm; LFR$\backslash$PLE is lens-forming region without presumptive lens ectoderm. Number is developmental stage. Percentage is lens formation rate. Question mark means the experiment is not executed.}
	\label{HG}
\end{table}

(3) In the experiments reported by Hamburger \cite{hamburger1939development}, chick limb bud primordia (LB) was transplanted to the right side of the chick body (CB) at the same developmental stage. The grafted tissues might lead to normal limbs (normal donor, ND), defective limbs (abnormal donor, AD), or atypical outgrowth (totally abnormal, TA). Tissues are from six developmental stages, directly named stage 1 to stage 6. The results are presented in Table \ref{Ham}. For example, the entry with host ``LB1'' means that if we transplant chick limb bud primordia at stage 1 to chick body at stage 1, there will be 36\% cases of normal development as donor, 36\% cases of abnormal development as donor, and 28\% cases of totally abnormal development.

\begin{table}[]
	\begin{tabular}{lllllllll}
		&                                                                          &  &         &         & \multicolumn{2}{l}{Donor tissue} &         &         \\
		&                                                                          &  & LB1     & LB2     & LB3             & LB4            & LB5     & LB6     \\
		&                                                                          &  &         &         &                 &                &         &         \\
		&                                                                          &  & ND 36\% & ND 58\% & ND 83\%         & ND 61\%        & ND 39\% & ND \, 9\%  \\
		\begin{tabular}[c]{@{}l@{}}Host\\ tissue\end{tabular} & \begin{tabular}[c]{@{}l@{}}CB at the same\\ stage with donor\end{tabular} &  & AD 36\% & AD 25\% & AD \, 4\%          & AD 13\%        & AD 17\% & AD 11\% \\
		&                                                                          &  & TA 28\% & TA 17\% & TA 13\%         & TA 26\%        & TA 44\% & TA 80\%
	\end{tabular}
	\caption{Transplantation results reported by Hamburger \cite{hamburger1939development}. LB is chick limb bud primordia, CB is right side of chick body. Number is developmental stage. Grafted tissues could develop normally as donor tissue (ND), develop abnormally as donor tissue (AD), or develop totally abnormally (TA), with different probabilities.}
	\label{Ham}
\end{table}

(4) In the experiments reported by Jones and Woodland \cite{jones1987development}, \emph{Xenopus laevis} animal caps at different stages were transplanted to \emph{Xenopus Borealis} vegetal plugs at different stages to observe whether the grafted tissue could be induced to form mesoderm. The experimental results are stochastic: some cases had induced mesoderm; some did not. The reformulated results are presented in Table \ref{Jo}. 

\begin{table}[]
	\begin{tabular}{lllllllllll}
		&       &     &        &        &        & \multicolumn{2}{l}{Donor tissue} &       &        &     \\
		&       & A4  & A5     & A6     & A7     & A8           & A9         & A10   & A10.5  & A11 \\
		& D5    & ?   & ?      & ?      & ?      & ?            & 82\%       & 82\%  & 62.5\% & ?   \\
		& D6    & ?   & ?      & ?      & ?      & ?            & ?          & 50\%  & 62.5\% & ?   \\
		& D7    & ?   & ?      & ?      & ?      & ?            & ?          & 45\%  & 20\%   & 0\% \\
		& D8    & ?   & ?      & ?      & ?      & ?            & ?          & 100\% & 20\%   & 0\% \\
		Host & D9    & ?   & 62.5\% & 85\%   & 85\%   & ?            & ?          & 100\% & ?      & ?   \\
		tissue& D10   & 8\% & 72\%   & 77\%   & 77\%   & ?            & ?          & 100\% & ?      & ?   \\
		& D10.5 & 0\% & 8\%    & 54.5\% & 54.5\% & 54.5\%       & ?          & ?     & ?      & ?   \\
		& D11   & ?   & ?      & ?      & ?      & 0\%          & 0\%        & 0\%   & ?      & ?   \\
		& D12   & ?   & ?      & ?      & ?      & 0\%          & 0\%        & 0\%   & ?      & ?  
	\end{tabular}
	\caption{Transplantation results reported by Jones and Woodland \cite{jones1987development} (reformulated). A4--A11 are animal caps of \emph{Xenopus laevis} at corresponding stages. D5--D12 are vegetal plugs of \emph{Xenopus borealis} at corresponding stages. Percentage is mesoderm induction rate. Question mark means the experiment is not executed.}
	\label{Jo}
\end{table}

(5) In the experiments reported by Arresta et al. \cite{arresta2005lens}, different \emph{Xenopus laevis} tissues were transplanted to the vitreous chamber of the right eye (lens removed) to check whether the grafted tissue could develop into lens tissue. The experimental results are stochastic: some cases had lens formation, some did not. The results are presented in Table \ref{AB}. 

\begin{table}[]
	\begin{tabular}{llllll}
		&      &       & \multicolumn{3}{l}{Donor tissue} \\
		&      & HE24  & HE26      & HE30      & HE40     \\
		Host tissue & VC55 & 60\%  & 47\%      & 33\%      & 27\%     \\
		&      &       &           &           &          \\
		&      &       & \multicolumn{3}{l}{Donor tissue} \\
		&      & VFE24 & VFE26     & VFE30     & VFE40    \\
		Host tissue & VC55 & 13\%  &  \, 7\%       &  \, 0\%       & \,  0\%      \\
		&      &       &           &           &          \\
		&      &       & \multicolumn{3}{l}{Donor tissue} \\
		&      & EVF44 & EVF46     & EVF48     &          \\
		Host tissue & VC55 & \,  0\%   &  \, 0\%       &  \, 0\%       &          \\
		&      &       &           &           &          \\
		&      &       & \multicolumn{3}{l}{Donor tissue} \\
		&      & EE44  & EE46      & EE48      &          \\
		Host tissue & VC55 & 20\%  & \,  5\%       &  \, 0\%       &         \\
		&&&&&
	\end{tabular}
	\caption{\emph{Xenopus laevis} transplantation results reported by Arresta et al. \cite{arresta2005lens}. HE is ectoderm above the forebrain; VFE is ventral part of flank ectoderm; EVF is epidermis of the ventral part of the flank; EE is epidermis above the forebrain; VC is vitreous chamber of right eye (lens removed). Number is developmental stage. Percentage is lens formation rate.}
	\label{AB}
\end{table}

(6) In the experiments reported by Elliott et al. \cite{elliott2013transplantation}, ear, heart, liver, or somite of \emph{Xenopus laevis} were transplanted to the orbit with eye removed. The experimental results are stochastic: some developed normally as donor tissue (ND), some developed abnormally as donor tissue (AD), some just died out (DE). The results are shown in Table \ref{EH}.

\begin{table}[]
	\begin{tabular}{llllll}
		&                  &             & \multicolumn{3}{l}{Donor tissue}  \\
		&                  & Ear,        & Heart,   & Liver,   & Somite,     \\
		&                  & stage 24-26 & stage 27 & stage 42 & stage 24-25 \\
		&&&&&\\
		   &     & ND 56\%     & ND 32\%  & ND 90\%  & ND 100\%    \\
			\begin{tabular}[c]{@{}l@{}}Host\\ tissue\end{tabular} & 	\begin{tabular}[c]{@{}l@{}}Orbit without\\ eye, stage 24-26\end{tabular} & AD 36\%     & AD 42\%  &  AD  \, 0\%        &   AD \,  \, 0\%          \\
		&                  & DE  \, 8\%      & DE 26\%  & DE 10\%  &   DE  \,  \, 0\%         \\
		&&&&&
	\end{tabular}
	\caption{\emph{Xenopus laevis} transplantation results reported by Elliott et al. \cite{elliott2013transplantation}. Grafted tissues could develop normally as donor tissue (ND), develop abnormally as donor tissue (AD), or die (DE), with different probabilities.}
	\label{EH}
\end{table}

(7) In the experiments reported by Kao and Chang \cite{kao1996homeotic,kao1997limb}, claw tissues of different crabs (\emph{Cancer gracilis}, \emph{Cancer productus}, \emph{Cancer anthonyi}, \emph{Cancer jordani}) were transplanted to autotomized stumps of the fourth walking leg. The grafted tissues might lead to normal legs (normal host, NH), abnormal legs (abnormal host, AH), or claws (normal donor, ND). Table \ref{Kao} presents partial results. 

\begin{table}[]
	\begin{tabular}{llllllll}
		&                                                                &  &                                                                   &                                                                    & \multicolumn{2}{l}{Donor tissue}                                                                                                        &                                                                   \\
		&                                                                &  & Dactyl                                                            & \begin{tabular}[c]{@{}l@{}}Dactyl\\ contralateral\end{tabular}     & Pollex                                                             & \begin{tabular}[c]{@{}l@{}}Pollex\\ contralateral\end{tabular}     & Ischium                                                           \\
		&                                                                &  &                                                                   &                                                                    &                                                                    &                                                                    &                                                                   \\
		\begin{tabular}[c]{@{}l@{}}Host\\ tissue\end{tabular} & \begin{tabular}[c]{@{}l@{}}Fourth\\ walking\\ leg\end{tabular} &  & \begin{tabular}[c]{@{}l@{}}NH 92\%\\ AH \, 0\%\\ ND \, 8\%\end{tabular} & \begin{tabular}[c]{@{}l@{}}NH 79\%\\ AH 14\%\\ ND \, 7\%\end{tabular} & \begin{tabular}[c]{@{}l@{}}NH 75\%\\ AH 25\%\\ ND \, 0\%\end{tabular} & \begin{tabular}[c]{@{}l@{}}NH 70\%\\ AH \, 0\%\\ ND 30\%\end{tabular} & \begin{tabular}[c]{@{}l@{}}NH 91\%\\ AH \, 0\%\\ ND \, 9\%\end{tabular}
	\end{tabular}
	\caption{Transplantation results (incomplete) reported by Kao and Chang \cite{kao1996homeotic}. Different claw tissues of \emph{Cancer gracilis} were transplanted to the autotomized stump of the fourth walking leg. The result can be a normal leg (NH), abnormal leg (AH) or claw (ND) with different probabilities.}
	\label{Kao}
\end{table}

(8) In the experiments reported by Smith and Slack \cite{smith1983dorsalization}, the dorsal marginal zone of \emph{Xenopus laevis} at stage 10 was transplanted to the ventral marginal zone at stage 10, and the grafted tissue developed abnormally as the donor tissue. On the other hand, the ventral marginal zone at stage 10 was transplanted to the dorsal marginal zone at stage 10, and the grafted tissue either developed normally as the host tissue or became totally abnormal.

\section{Inference of the unknown experimental results}
\label{S3}
\subsection{Possible ideas on experimental results inference}
It is difficult to infer the unknown experimental results directly from the known experimental results. An empirical law summarized from one group of experiments could be falsified by another group of experiments. For example, from the experiments reported by Krneta-Stankic et al. \cite{krneta2010temporal}, one might guess that exchanging donor and host does not affect the result. However, this is not true in the experiments reported by Smith and Slack \cite{smith1983dorsalization}. The experiments reported by Arresta et al. \cite{arresta2005lens} imply that the normal development rate decreases as the developmental stage increases, which is not the case in the experiments reported by Henry and Grainger \cite{henry1987inductive}.

The experimental results can be represented by a matrix with unknown entries. This is similar to the ``matrix completion'' problem \cite{candes2010power,nguyen2019low}. The most common setting of matrix completion problems is: for an $n\times n$ matrix, only some entries are known. The goal is to find a matrix $M$, whose rank is at most $r$ ($r\ll n$), and minimizes a penalty function. Generally, we get a penalty if (1) entries of $M$ do not match our knowledge; (2) the norm of $M$ is large. Nevertheless, known methods deal with numerical matrices, not nominal matrices in our case. Besides, entries in our case cannot be added or multiplied, making ``rank'' not applicable. Therefore, methods for matrix completion problems are not suitable.

Still, introducing a penalty function is a good idea to evaluate the guesses of the unknown experimental results. The question is when we should apply a penalty. We have two basic observations: (1) if donor tissue and host tissue are similar/not similar, the transplantation result tends to be normal/abnormal; (2) similar experiments (with similar donors and similar hosts) tend to have similar results. Now the task is to clarify the similarity between tissues and design a proper penalty function.

\subsection{Similarities between tissues and between experiments}
\label{sim}
We expect that biological knowledge of similarities between tissues could provide partial prior knowledge on the unknown experimental results. For example, we can measure the similarity by comparing the transcriptome information or concentrations of some critical molecules between tissues. Another choice is to calculate the distance between tissues on the developmental tree \cite{wang2020model}. Nevertheless, the developmental history of tissues is highly tangled, and the distance between tissues is difficult to define \cite{bernardini2013atlas}.

In this paper, the aim is not to quantify the tissue similarity through experiments, but to illustrate what inference we can make when the tissue similarity has been given. Therefore, we artificially and rather arbitrarily assign the similarities between tissues and between experiments. We will also study the influence of different versions of similarities on the inference results. If we have enough data, we can use the inferred experimental results to check whether the assigned similarities are proper or not.

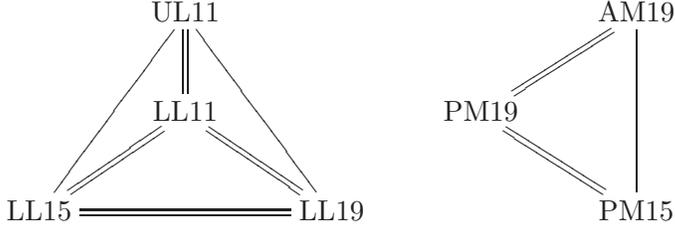
\begin{figure}
$\xymatrix{
	&\text{UL11}\ar@{=}[d]\ar@{-}[ddl]\ar@{-}[ddr]&&&\text{AM19}\ar@{-}[dd]\\
	&\text{LL11}\ar@{=}[dl]\ar@{=}[dr]&&\text{PM19}\ar@{=}[ur]\ar@{=}[dr]&\\
	\text{LL15}\ar@{=}[rr]&&\text{LL19}&&\text{PM15}
}$\\
\caption{Tissue similarity chart A for the experiments reported by Krneta-Stankic et al. \cite{krneta2010temporal}. Double/single/no line corresponds to high/medium/low similarity. AM is anterior paraxial mesoderm; PM is presomitic mesoderm; UL is upper lateral lip; LL is lower lip. Number is developmental stage.}
\label{KSF1}
\end{figure}

Consider the experiments reported by Krneta-Stankic et al. \cite{krneta2010temporal} in Table \ref{KS} with seven different tissues of two classes: mesoderms (AM19, PM19, PM15) and lips (UL11, LL11, LL15, LL19). We stipulate that a mesoderm tissue and a lip tissue (e.g., AM19 and LL11) have low similarity; the same type of tissue at different stages (e.g., LL11 and LL15) or similar types of tissues at the same stage (e.g., UL11 and LL11) have high similarity; other pairs (e.g., UL11 and LL15) have medium similarity. Figure \ref{KSF1} illustrates similarities between tissues. This version of the similarity relationship is named ``tissue similarity chart A'', and more versions will be discussed later. For each experiment, we can make a prediction: if the donor and the host have high or medium similarity, the result tends to be normal; otherwise, the result tends to be abnormal.

With the similarities between tissues being established, we can correspondingly define the similarities between experiments. Here each experiment is denoted by its donor and host. For example, \{AM19,LL11\} means the transplantation experiment with donor tissue AM19 and host tissue LL11. We stipulate that two experiments have high similarity if they have the same host and highly similar donors, or the same donor and highly similar hosts (e.g., \{UL11,LL11\} and \{UL11,LL15\}); two experiments have medium similarity if they have highly similar hosts and highly similar donors (e.g., \{AM19,UL11\} and \{PM19,LL11\}); other experiments have low similarity (e.g., \{AM19,UL11\} and \{PM15,LL15\}). 

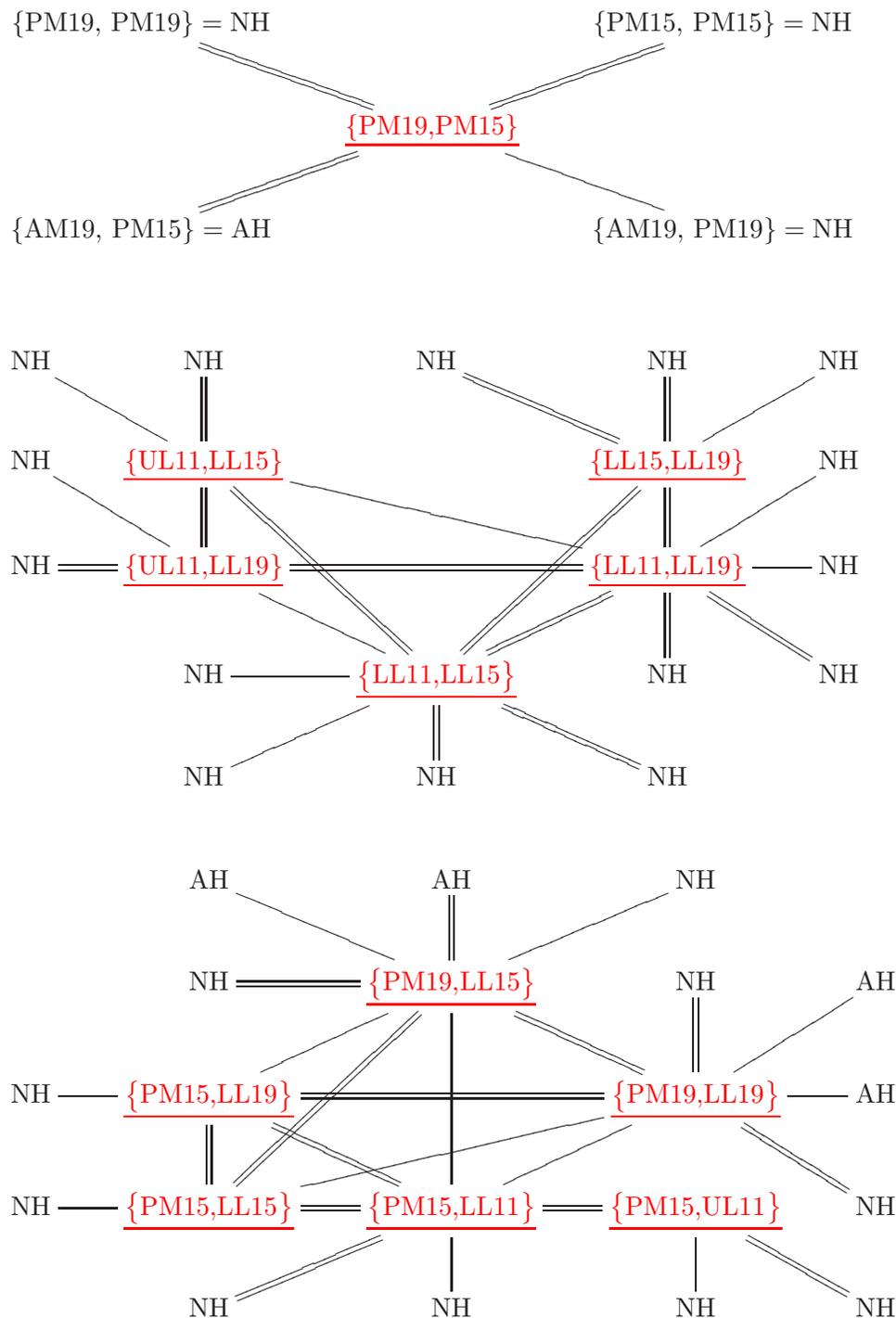
\begin{figure}
$\xymatrix{
	\{\text{PM19, PM19}\}=\text{NH} &             & \{\text{PM15, PM15}\}=\text{NH} \\
	&\color{red}{\underline{\{\text{PM19,PM15}\}}} \ar@{=}[lu]\ar@{=}[ru]\ar@{=}[ld]\ar@{-}[rd] &   \\
	\{\text{AM19, PM15}\}=\text{AH} &             & \{\text{AM19, PM19}\}=\text{NH}
}$\\

$\xymatrix{
	&&&&\\
	\text{NH} & \text{NH}           & \text{NH}           & \text{NH}           & \text{NH} \\
	\text{NH} & \color{red}{\underline{\{\text{UL11,LL15}\}}} \ar@{=}[d]\ar@{=}[ddr]\ar@{-}[drr] \ar@{-}[lu]\ar@{=}[u]&             &\color{red}{\underline{\{\text{LL15,LL19}\}}}\ar@{=}[ldd] \ar@{=}[u]\ar@{=}[ul]\ar@{-}[ur]& \text{NH} \\
	\text{NH} & \color{red}{\underline{\{\text{UL11,LL19}\}}} \ar@{=}[rr]\ar@{-}[lu]\ar@{=}[l] \ar@{-}[rd]&          & \color{red}{\underline{\{\text{LL11,LL19}\}}}  \ar@{=}[u]\ar@{=}[ld]\ar@{-}[ru]\ar@{-}[r]\ar@{=}[rd]\ar@{=}[d]& \text{NH} \\
	& \text{NH}           &\color{red}{\underline{{\begin{Bmatrix} \text{LL11,} \text{LL15} \end{Bmatrix}}}}\ar@{-}[l]\ar@{-}[ld]\ar@{=}[d]\ar@{=}[rd] & \text{NH}           & \text{NH} \\
	& \text{NH}           & \text{NH}           & \text{NH}           &  
}$\\
$\xymatrix{
	&&&&\\
	& \text{AH}  \ar@{-}[rd]        & \text{AH}    \ar@{=}[d]      & \text{NH}   \ar@{-}[ld]        &    \\
	& \text{NH}     \ar@{=}[r]      & \color{red}{\underline{{\begin{Bmatrix} \text{PM19,} \text{LL15} \end{Bmatrix}}}}  \ar@{-}[ld]\ar@{=}[ldd]\ar@{-}[dd]\ar@{=}[rd]& \text{NH}     \ar@{=}[d]      & \text{AH}\ar@{-}[ld] \\
	\text{NH}\ar@{-}[r] & \color{red}{\underline{{\begin{Bmatrix} \text{PM15,LL19} \end{Bmatrix}}}}  \ar@{=}[d]\ar@{=}[rd]\ar@{=}[rr]&             &\color{red}{\underline{{\begin{Bmatrix} \text{PM19,LL19} \end{Bmatrix}}}} & \text{AH}\ar@{-}[l] \\
	\text{NH} \ar@{-}[r]&\color{red}{\underline{{\begin{Bmatrix} \text{PM15,LL15} \end{Bmatrix}}}}  \ar@{-}[rru]\ar@{=}[r]&    \color{red}{\underline{{\begin{Bmatrix} \text{PM15,LL11} \end{Bmatrix}}}}    \ar@{=}[r]\ar@{-}[ru]       &\color{red}{\underline{{\begin{Bmatrix} \text{PM15,UL11} \end{Bmatrix}}}}  & \text{NH} \ar@{=}[lu] \\
	&     \text{NH}   \ar@{=}[ru]          & \text{NH}   \ar@{-}[u]  & \text{NH}   \ar@{-}[u]        & \text{NH}\ar@{=}[lu]  
}$\\
\caption{Similarities between experiments reported by Krneta-Stankic et al. \cite{krneta2010temporal}. \color{red}\underline{Red underlined}\color{black}/black entries are experiments with unknown/known results. Double/single/no line corresponds to high/medium/low similarity between experiments. To simplify the graph structure, the same experiment with known results can appear multiple times, and the similarities between experiments with known results are omitted. AM is anterior paraxial mesoderm; PM is presomitic mesoderm; UL is upper lateral lip; LL is lower lip. Number is developmental stage. NH means normal host; AH means abnormal host.}
\label{KSF2}
\end{figure}

To simplify the problem, we assume that exchanging donor and host does not affect the experimental result (as shown in the experiments reported by Smith and Slack \cite{smith1983dorsalization}, this is not always true), and they are regarded as the same experiment. Due to such symmetry, six results can be presumed (all with donor AM19). Besides, four experiments on the diagonal, namely those with the same host and donor (e.g., \{LL19,LL19\}), were not executed. Since it is the transplantation of one tissue to itself, the result can be presumed to be NH. All these presumed results are in \emph{italic} font in corresponding tables. With these presumptions, the number of distinct experiments with unknown results is reduced to 12. Figure \ref{KSF2} illustrates similarities between experiments, determined by tissue similarity chart A (Figure \ref{KSF1}). Similar experiments tend to have the same results.

\subsection{Penalty function}
We have constructed a graph where nodes are experiments, and edges describe similarities between experiments. Each experiment has two possible results, normal (+1) and abnormal (-1). We need a penalty function so that a configuration of guesses on unknown experimental results gets penalized if (1) similar experiments have different results; (2) the result of an experiment violates our prediction. 

We can refer to the Ising model \cite{cipra1987introduction} in ferromagnetism, which allows phase transition. It considers a set of lattice sites (e.g., 2D square lattice), where each site $k$ has a variable $\sigma_k$ that takes $+1$ or $-1$. For each pair of neighboring sites $i,j$, there is a coefficient $J_{ij}\ge 0$ that describes the interaction between $i,j$. For each site $j$, there is a coefficient $h_j$ that represents the external field. For a configuration $\sigma$ of variables over all sites, the energy is given by 
\[\mathrm{H}(\sigma)=-\sum_{i\sim j}J_{ij}\sigma_i\sigma_j-\sum_j h_j\sigma_j,\]
where $i\sim j$ means sites $i,j$ are neighboring. The probability of a configuration $\sigma$ is 
\[\mathbb{P}_\beta(\sigma)=e^{-\beta \mathrm{H}(\sigma)}/Z_\beta,\]
where $\beta=(k_B T)^{-1}$, and $Z_\beta=\sum_{\sigma}e^{-\beta \mathrm{H}(\sigma)}$ is the normalization constant. Configuration with high energy (high penalty) has small probability. Therefore, a configuration is less likely (with high penalty) if (1) neighboring sites have different values; (2) the value $\sigma_j$ and the external field $h_j$ have different signs (incompatible). 

Now we can see the analogy between this model and tissue transplantation: lattice$\Leftrightarrow$graph; site$\Leftrightarrow$experiment; binary variable ($+1,-1$)$\Leftrightarrow$result (normal, abnormal); neighboring sites$\Leftrightarrow$similar experiments; external field$\Leftrightarrow$prediction. The penalty conditions also have analogies: (1) neighboring sites ($\Leftrightarrow$similar experiments) have different values ($\Leftrightarrow$results); (2) site value ($\Leftrightarrow$result) and external field ($\Leftrightarrow$prediction) are incompatible.

To this point, the final analogy emerges: energy function$\Leftrightarrow$penalty function. It is clear that we can use the energy function $\mathrm{H}(\sigma)$ as our penalty function.

We need to warn that the analogy does not mean any physical relationship between tissue transplantation and ferromagnetism or phase transition. Also, the parameters we shall use (especially $\beta$) do not have physical meanings. 

In this paper, we slightly modify the external field term $h_j\sigma_j$, and adopt the following form of penalty function:
\[\mathrm{H}(\sigma)=-\sum_{i\sim j}J_{ij}\sigma_i\sigma_j-\sum_j h_j \pi_j\sigma_j.\]
Here $\sigma_i$ is the result of experiment $i$, taking value $+1$ or $-1$; $J_{ij}$ describes the strength of similarity between experiments $i,j$; $h_j\ge 0$ describes the strength of prediction; $\pi_j$ is the prediction of experiment $j$, taking value $+1$ or $-1$ (if we do not have a prediction, ignore $\pi_j$ and set $h_j=0$).

In the experiments reported by Krneta-Stankic et al. \cite{krneta2010temporal}, regard the result NH as $+1$, and AH as $-1$. For any experiment, set $h_j=h_0$, where $h_0$ is a properly chosen parameter. For an experiment $j$ that donor and host have high or medium similarity, set $\pi_j=1$; otherwise set $\pi_j=-1$. For two experiments $i,j$ that have high similarity, set $J_{ij}=2J_0$; for medium similarity, set $J_{ij}=J_0$; otherwise, set $J_{ij}=0$. Here $J_0$ is a properly chosen parameter. For a configuration $\{\sigma_i\}$, we can calculate its penalty function, and define its probability as:
$\mathbb{P}_\beta(\sigma)=e^{-\beta \mathrm{H}(\sigma)}/Z_\beta,$
where $\beta$ is a properly chosen parameter, and $Z_\beta=\sum_{\sigma}e^{-\beta \mathrm{H}(\sigma)}$ is the normalization constant.

\subsection{Inference results under different conditions}
For the experiments reported by Krneta-Stankic et al. \cite{krneta2010temporal}, we can calculate the probability $\mathbb{P}_\beta(\sigma)$ of each configuration $\sigma$ with chosen values of parameters $J_0,h_0,\beta$. We can determine the most probable configuration and calculate the expectation of all configurations (the percentage of each experiment to be NH or AH). 

For experiment similarities determined by tissue similarity chart A (Figure \ref{KSF1}), Tables \ref{KT1},\ref{KT2},\ref{KT3} present the most probable configuration for different values of $J_0,h_0$. Notice that $\beta$ does not affect which the most probable configuration is. Tables \ref{KT4},\ref{KT5},\ref{KT6},\ref{KT7} present the expectation of all configurations (in the form of NH percentage) for different values of $\beta,J_0,h_0$. \color{red}{\underline{Red underlined}} \color{black} entries are inferred results, \emph{black italic} entries are presumed results , and black normal entries are reported results. Under different choices of parameter values, the inferred results keep being reasonable.

\begin{table}[]
	\begin{tabular}{lllllllll}
		&    &    &    &    & \multicolumn{4}{l}{Donor} \\
		&    & AM19 & PM19 & PM15 & UL11   & LL11   & LL15   & LL19   \\
		& AM19 & \emph{NH}  & NH  & AH & AH   & AH   & AH   & NH    \\
		& PM19 & \emph{NH}  & NH  & \color{red}{\underline{NH}}  & NH    & NH    & \color{red}{\underline{NH}}   & \color{red}{\underline{NH}}   \\
		& PM15 & \emph{AH} & \color{red}{\underline{NH}}  & \emph{NH}  & \color{red}{\underline{NH}}   & \color{red}{\underline{NH}}   & \color{red}{\underline{NH}}   & \color{red}{\underline{NH}}   \\
		Host &UL11 & \emph{AH} & NH  & \color{red}{\underline{NH}} & NH    & NH    & \color{red}{\underline{NH}}    & \color{red}{\underline{NH}}    \\
		& LL11 & \emph{AH} & NH  & \color{red}{\underline{NH}} & NH    & NH    & \color{red}{\underline{NH}}    & \color{red}{\underline{NH}}    \\
		& LL15 & \emph{AH} & \color{red}{\underline{NH}} & \color{red}{\underline{NH}} & \color{red}{\underline{NH}}    & \color{red}{\underline{NH}}    & \emph{NH}    & \color{red}{\underline{NH}}    \\
		& LL19 & \emph{NH}  & \color{red}{\underline{NH}} & \color{red}{\underline{NH}} & \color{red}{\underline{NH}}    & \color{red}{\underline{NH}}    & \color{red}{\underline{NH}}    & \emph{NH}  \\
		&&&&&&&&  
	\end{tabular}
	\caption{Inferred most probable configuration in the experiments reported by Krneta-Stankic et al. \cite{krneta2010temporal}, with $J_0=1$, $h_0=1$ and tissue similarity chart A. \color{red}{\underline{Red underlined}} \color{black} entries are inferred results, \emph{black italic} entries are presumed results, and black normal entries are reported results. The value of $\beta$ does not affect in determining the most probable configuration.}
	\label{KT1}
\end{table}

\begin{table}[]
	\begin{tabular}{lllllllll}
		&    &    &    &    & \multicolumn{4}{l}{Donor} \\
		&    & AM19 & PM19 & PM15 & UL11   & LL11   & LL15   & LL19   \\
		& AM19 & \emph{NH}  & NH  & AH & AH   & AH   & AH   & NH    \\
		& PM19 & \emph{NH}  & NH  & \color{red}{\underline{NH}}  & NH    & NH    & \color{red}{\underline{AH}}   & \color{red}{\underline{AH}}   \\
		 & PM15 & \emph{AH} & \color{red}{\underline{NH}}  & \emph{NH}  & \color{red}{\underline{AH}}   & \color{red}{\underline{AH}}   & \color{red}{\underline{AH}}   & \color{red}{\underline{AH}}   \\
		Host &UL11 & \emph{AH} & NH  & \color{red}{\underline{AH}} & NH    & NH    & \color{red}{\underline{NH}}    & \color{red}{\underline{NH}}    \\
		 & LL11 & \emph{AH} & NH  & \color{red}{\underline{AH}} & NH    & NH    & \color{red}{\underline{NH}}    & \color{red}{\underline{NH}}    \\
		 & LL15 & \emph{AH} & \color{red}{\underline{AH}} & \color{red}{\underline{AH}} & \color{red}{\underline{NH}}    & \color{red}{\underline{NH}}    & \emph{NH}    & \color{red}{\underline{NH}}    \\
		& LL19 & \emph{NH}  & \color{red}{\underline{AH}} & \color{red}{\underline{AH}} & \color{red}{\underline{NH}}    & \color{red}{\underline{NH}}    & \color{red}{\underline{NH}}    & \emph{NH}    \\
		&&&&&&&&
	\end{tabular}
	\caption{Inferred most probable configuration in the experiments reported by Krneta-Stankic et al. \cite{krneta2010temporal}, with $J_0=0.5$, $h_0=1$ and tissue similarity chart A. \color{red}{\underline{Red underlined}} \color{black} entries are inferred results, \emph{black italic} entries are presumed results, and black normal entries are reported results. The value of $\beta$ does not affect in determining the most probable configuration.}
	\label{KT2}
\end{table}

\begin{table}[]
	\begin{tabular}{lllllllll}
		&    &    &    &    & \multicolumn{4}{l}{Donor} \\
		&    & AM19 & PM19 & PM15 & UL11   & LL11   & LL15   & LL19   \\
		& AM19 & \emph{NH}  & NH  & AH & AH   & AH   & AH   & NH    \\
		& PM19 & \emph{NH}  & NH  & \color{red}{\underline{NH}}  & NH    & NH    & \color{red}{\underline{NH}}   & \color{red}{\underline{NH}}   \\
		& PM15 & \emph{AH} & \color{red}{\underline{NH}}  & \emph{NH}  & \color{red}{\underline{NH}}   & \color{red}{\underline{NH}}   & \color{red}{\underline{NH}}   & \color{red}{\underline{NH}}   \\
		Host &UL11 & \emph{AH} & NH  & \color{red}{\underline{NH}} & NH    & NH    & \color{red}{\underline{NH}}    & \color{red}{\underline{NH}}    \\
		& LL11 & \emph{AH} & NH  & \color{red}{\underline{NH}} & NH    & NH    & \color{red}{\underline{NH}}    & \color{red}{\underline{NH}}    \\
		& LL15 & \emph{AH} & \color{red}{\underline{NH}} & \color{red}{\underline{NH}} & \color{red}{\underline{NH}}    & \color{red}{\underline{NH}}    & \emph{NH}    & \color{red}{\underline{NH}}    \\
		& LL19 & \emph{NH}  & \color{red}{\underline{NH}} & \color{red}{\underline{NH}} & \color{red}{\underline{NH}}    & \color{red}{\underline{NH}}    & \color{red}{\underline{NH}}    & \emph{NH}  \\
		&&&&&&&&  
	\end{tabular}
	\caption{Inferred most probable configuration in the experiments reported by Krneta-Stankic et al. \cite{krneta2010temporal}, with $J_0=1$, $h_0=0.5$ and tissue similarity chart A. \color{red}{\underline{Red underlined}} \color{black} entries are inferred results, \emph{black italic} entries are presumed results, and black normal entries are reported results. The value of $\beta$ does not affect in determining the most probable configuration.}
	\label{KT3}
\end{table}

\begin{table}[]
	\begin{tabular}{lllllllll}
		&    &    &    &    & \multicolumn{4}{l}{Donor} \\
		&    & AM19 & PM19 & PM15 & UL11   & LL11   & LL15   & LL19   \\
		& AM19 & \emph{100\%}  & 100\%  & 0\% & 0\%   & 0\%   & 0\%   & 100\%    \\
		& PM19 & \emph{100\%}  & 100\%     & \color{red}{\underline{65\%}}  & 100\%    & 100\%    & \color{red}{\underline{49\%}} & \color{red}{\underline{56\%}} \\
		& PM15 & \emph{0\%} & \color{red}{\underline{65\%}}  & \emph{100\%}     & \color{red}{\underline{62\%}} & \color{red}{\underline{62\%}} & \color{red}{\underline{53\%}} & \color{red}{\underline{54\%}} \\
		Host &UL11 & \emph{0\%} & 100\%     & \color{red}{\underline{62\%}}  & 100\%    & 100\%    & \color{red}{\underline{81\%}}  & \color{red}{\underline{81\%}}  \\
		& LL11 & \emph{0\%} & 100\%     & \color{red}{\underline{62\%}}  & 100\%    & 100\%    & \color{red}{\underline{90\%}}  & \color{red}{\underline{90\%}}  \\
		& LL15 & \emph{0\%} & \color{red}{\underline{49\%}} & \color{red}{\underline{53\%}} & \color{red}{\underline{81\%}} & \color{red}{\underline{90\%}} & \emph{100\%}     & \color{red}{\underline{86\%}}  \\
		& LL19 & \emph{100\%}  & \color{red}{\underline{56\%}} & \color{red}{\underline{54\%}} & \color{red}{\underline{81\%}} & \color{red}{\underline{90\%}} & \color{red}{\underline{86\%}}  & \emph{100\%}   \\
		&&&&&&&&
	\end{tabular}
	\caption{Inferred probability of having ``NH'' result in the experiments reported by Krneta-Stankic et al. \cite{krneta2010temporal}. \color{red}{\underline{Red underlined}} \color{black} entries are inferred results, \emph{black italic} entries are presumed results, and black normal entries are reported results. Calculated by taking expectations over all configurations, with $\beta=0.1$, $J_0=1$, $h_0=1$ and tissue similarity chart A.}
	\label{KT4}
\end{table}

\begin{table}[]
	\begin{tabular}{lllllllll}
		&    &    &    &    & \multicolumn{4}{l}{Donor} \\
		&    & AM19 & PM19 & PM15 & UL11   & LL11   & LL15   & LL19   \\
		& AM19 & \emph{100\%}  & 100\%  & 0\% & 0\%   & 0\%   & 0\%   & 100\%    \\
		& PM19 & \emph{100\%}  & 100\%     & \color{red}{\underline{57\%}}  & 100\%    & 100\%    & \color{red}{\underline{45\%}} & \color{red}{\underline{49\%}} \\
		 & PM15 & \emph{0\%} & \color{red}{\underline{57\%}}  & \emph{100\%}     & \color{red}{\underline{53\%}} & \color{red}{\underline{52\%}} & \color{red}{\underline{47\%}} & \color{red}{\underline{47\%}} \\
		Host &UL11 & \emph{0\%} & 100\%     & \color{red}{\underline{52\%}}  & 100\%    & 100\%    & \color{red}{\underline{67\%}}  & \color{red}{\underline{67\%}}  \\
		 & LL11 & \emph{0\%} & 100\%     & \color{red}{\underline{52\%}}  & 100\%    & 100\%    & \color{red}{\underline{74\%}}  & \color{red}{\underline{74\%}}  \\
		 & LL15 & \emph{0\%} & \color{red}{\underline{45\%}} & \color{red}{\underline{47\%}} & \color{red}{\underline{67\%}} & \color{red}{\underline{74\%}} & \emph{100\%}     & \color{red}{\underline{71\%}}  \\
		& LL19 & \emph{100\%}  & \color{red}{\underline{49\%}} & \color{red}{\underline{47\%}} & \color{red}{\underline{67\%}} & \color{red}{\underline{74\%}} & \color{red}{\underline{71\%}}  & \emph{100\%}   \\
		&&&&&&&&
	\end{tabular}
	\caption{Inferred probability of having ``NH'' result in the experiments reported by Krneta-Stankic et al. \cite{krneta2010temporal}. \color{red}{\underline{Red underlined}} \color{black} entries are inferred results, \emph{black italic} entries are presumed results, and black normal entries are reported results. Calculated by taking expectations over all configurations, with $\beta=0.1$, $J_0=0.5$, $h_0=1$ and tissue similarity chart A.}
	\label{KT5}
\end{table}

\begin{table}[]
	\begin{tabular}{lllllllll}
		&    &    &    &    & \multicolumn{4}{l}{Donor} \\
		&    & AM19 & PM19 & PM15 & UL11   & LL11   & LL15   & LL19   \\
		& AM19 & \emph{100\%}  & 100\%  & 0\% & 0\%   & 0\%   & 0\%   & 100\%    \\
		& PM19 & \emph{100\%}  & 100\%     & \color{red}{\underline{65\%}}  & 100\%    & 100\%    & \color{red}{\underline{54\%}} & \color{red}{\underline{61\%}} \\
		& PM15 & \emph{0\%} & \color{red}{\underline{65\%}}  & \emph{100\%}     & \color{red}{\underline{65\%}} & \color{red}{\underline{67\%}} & \color{red}{\underline{58\%}} & \color{red}{\underline{59\%}} \\
		Host &UL11 & \emph{0\%} & 100\%     & \color{red}{\underline{65\%}}  & 100\%    & 100\%    & \color{red}{\underline{79\%}}  & \color{red}{\underline{79\%}}  \\
		& LL11 & \emph{0\%} & 100\%     & \color{red}{\underline{67\%}}  & 100\%    & 100\%    & \color{red}{\underline{89\%}}  & \color{red}{\underline{89\%}}  \\
		& LL15 & \emph{0\%} & \color{red}{\underline{54\%}} & \color{red}{\underline{58\%}} & \color{red}{\underline{79\%}} & \color{red}{\underline{89\%}} & \emph{100\%}     & \color{red}{\underline{84\%}}  \\
		& LL19 & \emph{100\%}  & \color{red}{\underline{61\%}} & \color{red}{\underline{59\%}} & \color{red}{\underline{79\%}} & \color{red}{\underline{89\%}} & \color{red}{\underline{84\%}}  & \emph{100\%}   \\
		&&&&&&&&
	\end{tabular}
	\caption{Inferred probability of having ``NH'' result in the experiments reported by Krneta-Stankic et al. \cite{krneta2010temporal}. \color{red}{\underline{Red underlined}} \color{black} entries are inferred results, \emph{black italic} entries are presumed results, and black normal entries are reported results. Calculated by taking expectations over all configurations, with $\beta=0.1$, $J_0=1$, $h_0=0.5$ and tissue similarity chart A.}
	\label{KT6}
\end{table}

\begin{table}[]
	\begin{tabular}{lllllllll}
		&    &    &    &    & \multicolumn{4}{l}{Donor} \\
		&    & AM19 & PM19 & PM15 & UL11   & LL11   & LL15   & LL19   \\
		& AM19 & \emph{100\%}  & 100\%  & 0\% & 0\%   & 0\%   & 0\%   & 100\%    \\
		& PM19 & \emph{100\%}  & 100\%     & \color{red}{\underline{77\%}}  & 100\%    & 100\%    & \color{red}{\underline{61\%}} & \color{red}{\underline{68\%}} \\
		& PM15 & \emph{0\%} & \color{red}{\underline{77\%}}  & \emph{100\%}     & \color{red}{\underline{75\%}} & \color{red}{\underline{75\%}} & \color{red}{\underline{66\%}} & \color{red}{\underline{67\%}} \\
		Host &UL11 & \emph{0\%} & 100\%     & \color{red}{\underline{75\%}}  & 100\%    & 100\%    & \color{red}{\underline{97\%}}  & \color{red}{\underline{97\%}}  \\
		& LL11 & \emph{0\%} & 100\%     & \color{red}{\underline{75\%}}  & 100\%    & 100\%    & \color{red}{\underline{100\%}}  & \color{red}{\underline{100\%}}  \\
		& LL15 & \emph{0\%} & \color{red}{\underline{61\%}} & \color{red}{\underline{66\%}} & \color{red}{\underline{97\%}} & \color{red}{\underline{100\%}} & \emph{100\%}     & \color{red}{\underline{98\%}}  \\
		& LL19 & \emph{100\%}  & \color{red}{\underline{68\%}} & \color{red}{\underline{67\%}} & \color{red}{\underline{97\%}} & \color{red}{\underline{100\%}} & \color{red}{\underline{98\%}}  & \emph{100\%}   \\
		&&&&&&&&
	\end{tabular}
	\caption{Inferred probability of having ``NH'' result in the experiments reported by Krneta-Stankic et al. \cite{krneta2010temporal}. \color{red}{\underline{Red underlined}} \color{black} entries are inferred results, \emph{black italic} entries are presumed results, and black normal entries are reported results. Calculated by taking expectations over all configurations, with $\beta=0.2$, $J_0=1$, $h_0=1$ and tissue similarity chart A.}
	\label{KT7}
\end{table}

\begin{figure}
	$\xymatrix{
		&\text{UL11}\ar@{-}[d]\ar@{-}[ddl]\ar@{-}[ddr]&&&\text{AM19}\ar@{-}[dd]\\
		&\text{LL11}\ar@{=}[dl]\ar@{=}[dr]&&\text{PM19}\ar@{-}[ur]\ar@{=}[dr]&\\
		\text{LL15}\ar@{=}[rr]&&\text{LL19}&&\text{PM15}
	}$\\
	\caption{Tissue similarity chart B for the experiments reported by Krneta-Stankic et al. \cite{krneta2010temporal}. High similarity corresponds to double line; medium similarity corresponds to single line; low similarity corresponds to no line. AM is anterior paraxial mesoderm; PM is presomitic mesoderm; UL is upper lateral lip; LL is lower lip. Number is developmental stage.}
	\label{KSF3}
\end{figure}
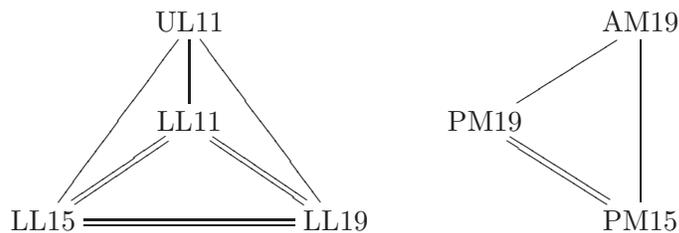

\begin{figure}
	$\xymatrix{
		&\text{UL11}\ar@{=}[d]\ar@{=}[ddl]\ar@{=}[ddr]&&&\text{AM19}\ar@{=}[dd]\\
		&\text{LL11}\ar@{=}[dl]\ar@{=}[dr]&&\text{PM19}\ar@{=}[ur]\ar@{=}[dr]&\\
		\text{LL15}\ar@{=}[rr]&&\text{LL19}&&\text{PM15}
	}$\\
	\caption{Tissue similarity chart C for the experiments reported by Krneta-Stankic et al. \cite{krneta2010temporal}. High similarity corresponds to double line; medium similarity corresponds to single line; low similarity corresponds to no line. AM is anterior paraxial mesoderm; PM is presomitic mesoderm; UL is upper lateral lip; LL is lower lip. Number is developmental stage.}
	\label{KSF4}
\end{figure}
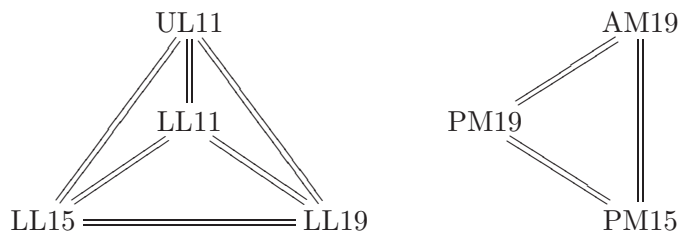

Besides the parameters, the tissue/experiment similarity relationship, shown as the structure of tissue/experiment similarity charts (Figures \ref{KSF1},\ref{KSF2}), also affects the inference results. We believe that the same tissue at different stages should still have high similarity, and one lip tissue and one mesoderm tissue should have low similarities. We consider two more tissue similarity charts (Figures \ref{KSF3},\ref{KSF4}), where we change the similarities between UL11 and LL11/LL15/LL19, and similarities between AM19 and PM15/PM19. With new tissue similarity charts, we use the same method in Section \ref{sim} to determine experiment similarities and use the same inference method with parameters for Tables \ref{KT1},\ref{KT4}. Figure \ref{KSF3} and Tables \ref{KTN1},\ref{KTN2} present tissue similarity chart B (experiment similarity chart omitted) and corresponding inference results. Figure \ref{KSF4} and Tables \ref{KTN3},\ref{KTN4} present tissue similarity chart C (experiment similarity chart omitted) and corresponding inference results. We can see that the change of tissue similarity chart (and thus the change of experiment similarity chart) has similar effects with the change of parameter values, and the inferred results are reasonable.

\begin{table}[]
	\begin{tabular}{lllllllll}
		&    &    &    &    & \multicolumn{4}{l}{Donor} \\
		&    & AM19 & PM19 & PM15 & UL11   & LL11   & LL15   & LL19   \\
		& AM19 & \emph{NH}  & NH  & AH & AH   & AH   & AH   & NH    \\
		& PM19 & \emph{NH}  & NH  & \color{red}{\underline{NH}}  & NH    & NH    & \color{red}{\underline{NH}}   & \color{red}{\underline{NH}}   \\
		& PM15 & \emph{AH} & \color{red}{\underline{NH}}  & \emph{NH}  & \color{red}{\underline{NH}}   & \color{red}{\underline{NH}}   & \color{red}{\underline{NH}}   & \color{red}{\underline{NH}}   \\
		Host &UL11 & \emph{AH} & NH  & \color{red}{\underline{NH}} & NH    & NH    & \color{red}{\underline{NH}}    & \color{red}{\underline{NH}}    \\
		& LL11 & \emph{AH} & NH  & \color{red}{\underline{NH}} & NH    & NH    & \color{red}{\underline{NH}}    & \color{red}{\underline{NH}}    \\
		& LL15 & \emph{AH} & \color{red}{\underline{NH}} & \color{red}{\underline{NH}} & \color{red}{\underline{NH}}    & \color{red}{\underline{NH}}    & \emph{NH}    & \color{red}{\underline{NH}}    \\
		& LL19 & \emph{NH}  & \color{red}{\underline{NH}} & \color{red}{\underline{NH}} & \color{red}{\underline{NH}}    & \color{red}{\underline{NH}}    & \color{red}{\underline{NH}}    & \emph{NH}  \\
		&&&&&&&&  
	\end{tabular}
	\caption{Inferred most probable configuration in the experiments reported by Krneta-Stankic et al. \cite{krneta2010temporal}, with $J_0=1$, $h_0=1$ and tissue similarity chart B. \color{red}{\underline{Red underlined}} \color{black} entries are inferred results, \emph{black italic} entries are presumed results, and black normal entries are reported results. The value of $\beta$ does not affect in determining the most probable configuration.}
	\label{KTN1}
\end{table}

\begin{table}[]
	\begin{tabular}{lllllllll}
		&    &    &    &    & \multicolumn{4}{l}{Donor} \\
		&    & AM19 & PM19 & PM15 & UL11   & LL11   & LL15   & LL19   \\
		& AM19 & \emph{100\%}  & 100\%  & 0\% & 0\%   & 0\%   & 0\%   & 100\%    \\
		& PM19 & \emph{100\%}  & 100\%     & \color{red}{\underline{69\%}}  & 100\%    & 100\%    & \color{red}{\underline{57\%}} & \color{red}{\underline{57\%}} \\
		& PM15 & \emph{0\%} & \color{red}{\underline{69\%}}  & \emph{100\%}     & \color{red}{\underline{55\%}} & \color{red}{\underline{57\%}} & \color{red}{\underline{54\%}} & \color{red}{\underline{54\%}} \\
		Host &UL11 & \emph{0\%} & 100\%     & \color{red}{\underline{55\%}}  & 100\%    & 100\%    & \color{red}{\underline{67\%}}  & \color{red}{\underline{67\%}}  \\
		& LL11 & \emph{0\%} & 100\%     & \color{red}{\underline{57\%}}  & 100\%    & 100\%    & \color{red}{\underline{84\%}}  & \color{red}{\underline{84\%}}  \\
		& LL15 & \emph{0\%} & \color{red}{\underline{57\%}} & \color{red}{\underline{54\%}} & \color{red}{\underline{67\%}} & \color{red}{\underline{84\%}} & \emph{100\%}     & \color{red}{\underline{84\%}}  \\
		& LL19 & \emph{100\%}  & \color{red}{\underline{57\%}} & \color{red}{\underline{54\%}} & \color{red}{\underline{67\%}} & \color{red}{\underline{84\%}} & \color{red}{\underline{84\%}}  & \emph{100\%}   \\
		&&&&&&&&
	\end{tabular}
	\caption{Inferred probability of having ``NH'' result in the experiments reported by Krneta-Stankic et al. \cite{krneta2010temporal}. \color{red}{\underline{Red underlined}} \color{black} entries are inferred results, \emph{black italic} entries are presumed results, and black normal entries are reported results. Calculated by taking expectations over all configurations, with $\beta=0.1$, $J_0=1$, $h_0=1$ and tissue similarity chart B.}
	\label{KTN2}
\end{table}

\begin{table}[]
	\begin{tabular}{lllllllll}
		&    &    &    &    & \multicolumn{4}{l}{Donor} \\
		&    & AM19 & PM19 & PM15 & UL11   & LL11   & LL15   & LL19   \\
		& AM19 & \emph{NH}  & NH  & AH & AH   & AH   & AH   & NH    \\
		& PM19 & \emph{NH}  & NH  & \color{red}{\underline{NH}}  & NH    & NH    & \color{red}{\underline{AH}}   & \color{red}{\underline{AH}}   \\
		& PM15 & \emph{AH} & \color{red}{\underline{NH}}  & \emph{NH}  & \color{red}{\underline{AH}}   & \color{red}{\underline{AH}}   & \color{red}{\underline{AH}}   & \color{red}{\underline{AH}}   \\
		Host &UL11 & \emph{AH} & NH  & \color{red}{\underline{AH}} & NH    & NH    & \color{red}{\underline{NH}}    & \color{red}{\underline{NH}}    \\
		& LL11 & \emph{AH} & NH  & \color{red}{\underline{AH}} & NH    & NH    & \color{red}{\underline{NH}}    & \color{red}{\underline{NH}}    \\
		& LL15 & \emph{AH} & \color{red}{\underline{AH}} & \color{red}{\underline{AH}} & \color{red}{\underline{NH}}    & \color{red}{\underline{NH}}    & \emph{NH}    & \color{red}{\underline{NH}}    \\
		& LL19 & \emph{NH}  & \color{red}{\underline{AH}} & \color{red}{\underline{AH}} & \color{red}{\underline{NH}}    & \color{red}{\underline{NH}}    & \color{red}{\underline{NH}}    & \emph{NH}    \\
		&&&&&&&&
	\end{tabular}
	\caption{Inferred most probable configuration in the experiments reported by Krneta-Stankic et al. \cite{krneta2010temporal}, with $J_0=1$, $h_0=1$ and tissue similarity chart C. \color{red}{\underline{Red underlined}} \color{black} entries are inferred results, \emph{black italic} entries are presumed results, and black normal entries are reported results. The value of $\beta$ does not affect in determining the most probable configuration.}
	\label{KTN3}
\end{table}

\begin{table}[]
	\begin{tabular}{lllllllll}
		&    &    &    &    & \multicolumn{4}{l}{Donor} \\
		&    & AM19 & PM19 & PM15 & UL11   & LL11   & LL15   & LL19   \\
		& AM19 & \emph{100\%}  & 100\%  & 0\% & 0\%   & 0\%   & 0\%   & 100\%    \\
		& PM19 & \emph{100\%}  & 100\%     & \color{red}{\underline{73\%}}  & 100\%    & 100\%    & \color{red}{\underline{47\%}} & \color{red}{\underline{55\%}} \\
		& PM15 & \emph{0\%} & \color{red}{\underline{73\%}}  & \emph{100\%}     & \color{red}{\underline{42\%}} & \color{red}{\underline{42\%}} & \color{red}{\underline{39\%}} & \color{red}{\underline{46\%}} \\
		Host &UL11 & \emph{0\%} & 100\%     & \color{red}{\underline{42\%}}  & 100\%    & 100\%    & \color{red}{\underline{95\%}}  & \color{red}{\underline{95\%}}  \\
		& LL11 & \emph{0\%} & 100\%     & \color{red}{\underline{42\%}}  & 100\%    & 100\%    & \color{red}{\underline{95\%}}  & \color{red}{\underline{95\%}}  \\
		& LL15 & \emph{0\%} & \color{red}{\underline{47\%}} & \color{red}{\underline{39\%}} & \color{red}{\underline{95\%}} & \color{red}{\underline{95\%}} & \emph{100\%}     & \color{red}{\underline{95\%}}  \\
		& LL19 & \emph{100\%}  & \color{red}{\underline{55\%}} & \color{red}{\underline{46\%}} & \color{red}{\underline{95\%}} & \color{red}{\underline{95\%}} & \color{red}{\underline{95\%}}  & \emph{100\%}   \\
		&&&&&&&&
	\end{tabular}
	\caption{Inferred probability of having ``NH'' result in the experiments reported by Krneta-Stankic et al. \cite{krneta2010temporal}. \color{red}{\underline{Red underlined}} \color{black} entries are inferred results, \emph{black italic} entries are presumed results, and black normal entries are reported results. Calculated by taking expectations over all configurations, with $\beta=0.1$, $J_0=1$, $h_0=1$ and tissue similarity chart C.}
	\label{KTN4}
\end{table}

\subsection{Workflow and remarks}
In summary, the procedure for inferring experiments with deterministic binary results is: (1) determine the similarities between tissues and between experiments, and related parameters; (2) for each configuration of unknown results, calculate its probability; (3) choose the most probable configuration, or take expectation on configurations. See Algorithm \ref{alg1} for the detailed workflow.

\begin{algorithm}[!htbp]
	\caption{Detailed workflow of the inference method for experiments with deterministic binary results.}
	\label{alg1}
	\vspace{-\bigskipamount}
	\begin{enumerate}
		\item \textbf{Input} \\
		\quad Results of some experiments (binary and deterministic)
		\item \textbf{Set} similarities between tissues\\
		\item \textbf{Set} similarities between experiments\\
		\item \textbf{Set} the values of coefficients\\
		\item {\bf For} each configuration $\sigma_0$ of unknown results,  \\
		\quad  {\bf Calculate} its penalty $\mathrm{H}(\sigma_0)$\\
		{\bf End} of for loop\\
		\item {\bf Calculate} the normalization constant $Z_\beta$\\
		\item {\bf For} each configuration $\sigma_0$ of unknown results,  \\
		\quad  {\bf Calculate} its probability $\mathbb{P}(\sigma_0)$\\
		{\bf End} of for loop\\
		\item {\bf Output} the most probable configuration
		\item {\bf Output} the expectation of all configurations
	\end{enumerate}
\end{algorithm}

Tissue/experiment similarities and parameters in the penalty function can affect the inference results. For the experiments reported by Krneta-Stankic et al. \cite{krneta2010temporal}, we show that adjusting each factor within a reasonable range does not prevent the inference results from being reasonable. These inference results (Tables \ref{KT1}-\ref{KTN4}) altogether prove that our method is robust under perturbations.

Our method compare the results of similar experiments. Therefore, if we have several similar experiments with unknown results, their results have to be inferred altogether as a group. Nevertheless, since the probability has an exponential form, if we have two groups of experiments with unknown results, and these two groups are separated by experiments with known results in the similarity chart, then the results of these two groups are independent, and can be inferred separately. For the experiments reported by Krneta-Stankic et al. \cite{krneta2010temporal}, 12 experiments with unknown results are separated into three groups, with 1, 5, 6 experiments (see Figure \ref{KSF2}). Thus we only need to consider $2^1=2$, $2^5=32$, $2^6=64$ configurations separately for different groups, not $2^{12}=4096$ configurations altogether.

The exponential form of $\mathbb{P}_\beta(\sigma)$ has another advantage: adding a constant to the penalty function $\mathrm{H}(\sigma)$ does not affect $\mathbb{P}_\beta(\sigma)$. Therefore, when calculating $\mathrm{H}(\sigma)$, we can omit some terms that are constants for all configurations, namely those terms that only concern experiments with known results.

When the tissue number is quite large, such that the normalization constant $Z_\beta$ is difficult to calculate, the expectation of configurations can be approximated by some Markov chain Monte Carlo methods, such as Glauber dynamics \cite{martinelli1994approach}.

\section{Inference for experiments with stochastic results}
\label{S4}

In Section \ref{S3}, we develop a method to conduct inference on experiments with deterministic binary results. In the experiments reported by Henry and Grainger \cite{henry1987inductive} in Table \ref{HG}, the experimental results are stochastic: we have percentages for lens formation (corresponds to ``normal'') and no lens formation (corresponds to ``abnormal''). We should not regard the percentage matrix as a numerical matrix and try matrix completion methods. Instead, we should sample deterministic configurations from stochastic results, and apply our method. 

For example, consider three similar experiments with results $[61\%\text{N}\text{ }? \text{ }58\%\text{N}]$. Here the first and the third experiments have 61\% and 58\% probabilities to be normal, and the second experimental result is unknown. We will first sample deterministic results for the first and the third experiments, and then use the method introduced in Section \ref{S3} to infer the result of the second experiment. The inferred result will be averaged over the samples of the first and the third experiments. When sampling deterministic results, we assume these experiments are independent. For example, the probability of sampling ``N'' and ``N'' for the first and the third experiments is $\mathbb{P}([\text{N ? N}])=61\%\times 58\%=35\%.$ Conditioned on this sample, we use the inference method to calculate the conditional probability for the second experiment to be normal, $\mathbb{P}(\text{?=N}\mid [\text{N ? N}])=98\%$. When we have considered all possible samples of the known (stochastic) experimental results, we can calculate the overall probability for the second experiment to be normal:
\[\mathbb{P}([\text{?=N}])=\mathbb{P}([\text{N ? N}])\times\mathbb{P}(\text{?=N}\mid [\text{N ? N}])+\mathbb{P}([\text{N ? A}])\times\mathbb{P}(\text{?=N}\mid [\text{N ? A}])\]
\[+\mathbb{P}([\text{A ? N}])\times\mathbb{P}(\text{?=N}\mid [\text{A ? N}])+\mathbb{P}([\text{A ? A}])\times\mathbb{P}(\text{?=N}\mid [\text{A ? A}])=59\%.\]

In general, denote the configuration of unknown experimental results as $\sigma$, and the configuration of known experimental results as $\rho$. For each configuration of known experimental results $\rho=\rho_0$, we can calculate its probability by assuming these experiments are independent: $\mathbb{P}(\rho=\rho_0)=\prod_i \mathbb{P}(\rho^i=\rho_0^i),$ as shown above. Then we apply the penalty function, and calculate the conditional expectation $\mathbb{E}(\sigma\mid \rho=\rho_0)$, same with the previous section. Last, take expectation with respect to $\rho$, to get the overall expectation of unknown experimental results $\mathbb{E}(\sigma)=\sum_{\rho_0}\mathbb{P}(\rho=\rho_0)\mathbb{E}(\sigma\mid \rho=\rho_0)$. See Algorithm \ref{alg2} for the detailed workflow.

\begin{algorithm}[!htbp]
	\caption{Detailed workflow of the inference method for experiments with stochastic binary results.}
	\label{alg2}
	\vspace{-\bigskipamount}
	\begin{enumerate}
		\item \textbf{Input} \\
		\quad Results of some experiments (binary and stochastic)\\
		\item \textbf{Set} similarities between tissues\\
		\item \textbf{Set} similarities between experiments\\
		\item \textbf{Set} the values of coefficients\\
		\item {\bf For} each configuration $\rho_0$ of known results\\
		\quad {\bf Calculate} its probability $\mathbb{P}(\rho=\rho_0)=\prod_i \mathbb{P}(\rho^i=\rho_0^i)$\\
		\quad {\bf For} each configuration $\sigma_0$ of unknown results,  \\
		\quad\quad  {\bf Calculate} its conditional penalty $\mathrm{H}(\sigma_0\mid \rho=\rho_0)$\\
		\quad {\bf End} of for loop\\
		\quad {\bf Calculate} the normalization constant $Z_\beta$\\
		\quad {\bf For} each configuration $\sigma_0$ of unknown results,  \\
		\quad\quad  {\bf Calculate} its conditional probability $\mathbb{P}(\sigma=\sigma_0\mid \rho=\rho_0)$\\
		\quad {\bf End} of for loop\\
		\quad {\bf Calculate} the conditional expectation $\mathbb{E}(\sigma\mid \rho=\rho_0)$\\
		{\bf End} of for loop\\
		\item {\bf Output} the overall expectation $\mathbb{E}(\sigma)=\sum_{\rho_0}\mathbb{P}(\rho=\rho_0)\mathbb{E}(\sigma\mid \rho=\rho_0)$
	\end{enumerate}
\end{algorithm}

We apply this method to the experiments reported by Henry and Grainger \cite{henry1987inductive}. Experiments that are neighboring in the table (e.g., \{PLE11,LFR$\backslash$PLE14\} and \{PLE12,LFR$\backslash$PLE14\}) have $J_{ij}=1$; otherwise set $J_{ij}=0$. For experiment \{PLE16,LFR$\backslash$PLE16\}, set $\pi_j=1$ and $h_j=1$; otherwise set $h_j=0$. Tables \ref{HG2},\ref{HG3} present the inferred probabilities (\color{red}{\underline{red underlined}}\color{black}) of lens formation under different values of parameter $\beta$. The results are reasonable for both values of $\beta$.

\begin{table}[]
	\begin{tabular}{lllllll}
		&                          &       &       & \multicolumn{3}{l}{Donor tissue} \\
		&                          & PLE11 & PLE12 & PLE14     & PLE16     & PLE19    \\
		& LFR\textbackslash{}PLE14 & 61\%  & 58\%  & 82\%      & \color{red}{\underline{93\%}}         & \color{red}{\underline{94\%}}        \\
		\begin{tabular}[c]{@{}l@{}}Host\\ tissue\end{tabular} & LFR\textbackslash{}PLE16 & \color{red}{\underline{39\%}}     & \color{red}{\underline{53\%}}     & \color{red}{\underline{88\%}}         & \color{red}{\underline{97\%}}         & \color{red}{\underline{97\%}}        \\
		& LFR\textbackslash{}PLE19 & 4\%   & 24\%  & 83\%      & \color{red}{\underline{96\%}}         & 100\%    \\
		&                          &       &       &           &           &          \\
		&                          &       &       & \multicolumn{3}{l}{Donor tissue} \\
		&                          & AVE11 & AVE12 & AVE14     & AVE16     & AVE19    \\
		& LFR\textbackslash{}PLE14 & 29\%  & 50\%  & 14\%      & 0\%       & 0\%      \\
		\begin{tabular}[c]{@{}l@{}}Host\\ tissue\end{tabular} & LFR\textbackslash{}PLE16 & \color{red}{\underline{9\%}}     & \color{red}{\underline{7\%}}     & \color{red}{\underline{1\%}}         & \color{red}{\underline{0\%}}         & \color{red}{\underline{0\%}}        \\
		& LFR\textbackslash{}PLE19 & 8\%   & 13\%  & 0\%       & \color{red}{\underline{0\%}}         & 0\%      \\
		&                          &       &       &           &           &          \\
		&                          &       &       & \multicolumn{3}{l}{Donor tissue} \\
		&                          & PVE11 & PVE12 & PVE14     & PVE16     & PVE19    \\
		\begin{tabular}[c]{@{}l@{}}Host\\ tissue\end{tabular} & LFR\textbackslash{}PLE14 & 11\%  & 16\%  & 4\%       & 0\%       & \color{red}{\underline{12\%}}    \\
		&&&&&&   
	\end{tabular}
	\caption{Inferred probability of lens formation in the experiments reported by Henry and Grainger \cite{henry1987inductive}. \color{red}{\underline{Red underlined}} \color{black} entries are inferred results, and black entries are reported results. Calculated by taking expectations over all configurations, with $\beta=1$.}
	\label{HG2}
\end{table}

\begin{table}[]
	\begin{tabular}{lllllll}
		&                          &       &       & \multicolumn{3}{l}{Donor tissue} \\
		&                          & PLE11 & PLE12 & PLE14     & PLE16     & PLE19    \\
		& LFR\textbackslash{}PLE14 & 61\%  & 58\%  & 82\%      & \color{red}{\underline{98\%}}         & \color{red}{\underline{98\%}}        \\
		\begin{tabular}[c]{@{}l@{}}Host\\ tissue\end{tabular} & LFR\textbackslash{}PLE16 & \color{red}{\underline{42\%}}     & \color{red}{\underline{56\%}}     & \color{red}{\underline{90\%}}         & \color{red}{\underline{98\%}}         & \color{red}{\underline{99\%}}        \\
		& LFR\textbackslash{}PLE19 & 4\%   & 24\%  & 83\%      & \color{red}{\underline{98\%}}         & 100\%    \\
		&                          &       &       &           &           &          \\
		&                          &       &       & \multicolumn{3}{l}{Donor tissue} \\
		&                          & AVE11 & AVE12 & AVE14     & AVE16     & AVE19    \\
		& LFR\textbackslash{}PLE14 & 29\%  & 50\%  & 14\%      & 0\%       & 0\%      \\
		\begin{tabular}[c]{@{}l@{}}Host\\ tissue\end{tabular} & LFR\textbackslash{}PLE16 & \color{red}{\underline{5\%}}     & \color{red}{\underline{5\%}}     & \color{red}{\underline{0\%}}         & \color{red}{\underline{0\%}}         & \color{red}{\underline{0\%}}        \\
		& LFR\textbackslash{}PLE19 & 8\%   & 13\%  & 0\%       & \color{red}{\underline{0\%}}         & 0\%      \\
		&                          &       &       &           &           &          \\
		&                          &       &       & \multicolumn{3}{l}{Donor tissue} \\
		&                          & PVE11 & PVE12 & PVE14     & PVE16     & PVE19    \\
		\begin{tabular}[c]{@{}l@{}}Host\\ tissue\end{tabular} & LFR\textbackslash{}PLE14 & 11\%  & 16\%  & 4\%       & 0\%       & \color{red}{\underline{2\%}}    \\
		&&&&&&   
	\end{tabular}
	\caption{Inferred probability of lens formation in the experiments reported by Henry and Grainger \cite{henry1987inductive}. \color{red}{\underline{Red underlined}} \color{black} entries are inferred results, and black entries are reported results. Calculated by taking expectations over all configurations, with $\beta=2$.}
	\label{HG3}
\end{table}

\section{Inference for experiments with multiple results}
\label{S5}
In the previous two sections, we only consider experiments with binary results. When there are at least three possible results, such as in the experiments reported by Hamburger \cite{hamburger1939development}, we need to modify the penalty function to describe predictions and similarities between experiments properly. There are many possible results for tissue transplantation experiments: transdifferentiation into a new type of normal tissue, transdifferentiation into a new type of abnormal tissue, normal development as the host, abnormal development as the host, normal development as the donor, abnormal development as the donor, totally abnormal development, death.

The penalty function for binary experiment is
\[\mathrm{H}(\sigma)=-\sum_{i\sim j}J_{ij}\sigma_i\sigma_j-\sum_j h_j \pi_j\sigma_j.\]
Here $\sigma_i\sigma_j=1$ for $\sigma_i=\sigma_j$, and $\sigma_i\sigma_j=-1$ for $\sigma_i=-\sigma_j$; $\pi_j\sigma_j=1$ for $\pi_j=\sigma_j$, and $\pi_j\sigma_j=-1$ for $\pi_j=-\sigma_j$. 

The product term $\sigma_i\sigma_j$ describes the similarity between results $\sigma_i$ and $\sigma_j$. We can replace this term by a comparison function between two results: $f(\sigma_i,\sigma_j)$. This function should be symmetric with two arguments $\sigma_i,\sigma_j$, and assign larger values for more similar results $\sigma_i,\sigma_j$. The term $\pi_j\sigma_j$ describes the similarity between the result $\sigma_j$ and the prediction $\pi_j$, and can be replaced by $f(\pi_j,\sigma_j)$. The new form of penalty function is
\[\mathrm{H}(\sigma)=-\sum_{i\sim j}J_{ij}f(\sigma_i,\sigma_j)-\sum_j h_j f(\pi_j,\sigma_j).\]
The probability of a configuration $\sigma$ is still $\mathbb{P}_\beta(\sigma)=e^{-\beta \mathrm{H}(\sigma)}/Z_\beta,$
where $\beta$ is a properly chosen parameter, and $Z_\beta=\sum_{\sigma}e^{-\beta \mathrm{H}(\sigma)}$ is the normalization constant. When the experiment has two possible results, the function returns to $f(\sigma_i,\sigma_j)=\sigma_i\sigma_j$ or its equivalent form.

For experiments with at least three possible results, the workflow is almost the same as Algorithm \ref{alg1} or Algorithm \ref{alg2}, except that we need one more step to define the comparison function $f$, and apply it in calculating the penalty.

We apply this method to the chick experiments reported by Hamburger \cite{hamburger1939development} in Table \ref{Ham}. The comparison function $f$ is defined as: $f(\text{ND},\text{ND})=f(\text{AD},\text{AD})=f(\text{TA},\text{TA})=2$, $f(\text{ND},\text{AD})=f(\text{TA},\text{AD})=0$, $f(\text{ND},\text{TA})=-1$, since from ND to AD to TA, the abnormality increases. For experiments with neighboring stages, set $J_{ij}=1$, otherwise set $J_{ij}=0$. Since there is no prior knowledge, set $h_j=0$ for all experiments.

These experiments were thoroughly conducted so that there is no unknown result for us to infer. Therefore we perform a ``cross validation'', meaning that we assign some experiments to be the training set, and use the results of the training set to infer other results (the testing set). Then we compare the inferred results and the real results on the testing set. Specifically, we apply the ``leave-one-out cross validation'', meaning that each time we choose one experiment to be the testing set, and use the other five experiments' results (training set) to conduct inference. Repeat this procedure for all six experiments, so that for each experiment, we can compare the inferred result and the real result. Since the experimental results are stochastic, we apply the mechanism discussed in Section \ref{S4}, namely choosing one configuration of known results randomly, conducting inference, then taking expectations.

Tables \ref{Ham2},\ref{Ham3} present the comparison between reported results (black) and inferred results (\color{red}{\underline{red underlined}}\color{black}). Our inference method produces satisfactory results for experiments at stages 2,4,5, under different values of parameter $\beta$. For experiments at stages 1,6, each one is only similar to one experiment with known results; thus there is not enough information to conduct reliable inferences. For the experiment at stage 3, the real results are maximum/minimum among all experiments, not similar to neighboring experiments 2,4. Theoretically speaking, we cannot predict such outlier cases without additional experimental information. 

\begin{table}[]
	\begin{tabular}{lllllllll}
		&                                                                           &  &                                                           &                                                           & \multicolumn{2}{l}{Donor tissue}                                                                                       &                                                            &                                                           \\
		&                                                                           &  & LB1                                                       & LB2                                                       & LB3                                                        & LB4                                                       & LB5                                                        & LB6                                                       \\
		&                                                                           &  &                                                           &                                                           &                                                            &                                                           &                                                            &                                                           \\
		&                                                                           &  & \begin{tabular}[c]{@{}l@{}}ND 36\%\\ \color{red}{\underline{ND 52\%}}\end{tabular} & \begin{tabular}[c]{@{}l@{}}ND 58\%\\ \color{red}{\underline{ND 57\%}}\end{tabular} & \begin{tabular}[c]{@{}l@{}}ND 83\%\\ \color{red}{\underline{ND 57\%}}\end{tabular}  & \begin{tabular}[c]{@{}l@{}}ND 61\%\\ \color{red}{\underline{ND 57\%}}\end{tabular} & \begin{tabular}[c]{@{}l@{}}ND 39\% \\ \color{red}{\underline{ND 31\%}}\end{tabular} & \begin{tabular}[c]{@{}l@{}}ND \, 9\%\\ \color{red}{\underline{ND 36\%}}\end{tabular}  \\
		&                                                                           &  &                                                           &                                                           &                                                            &                                                           &                                                            &                                                           \\
		\begin{tabular}[c]{@{}l@{}}Host\\ tissue\end{tabular} & \begin{tabular}[c]{@{}l@{}}CB at the same\\ stage with donor\end{tabular} &  & \begin{tabular}[c]{@{}l@{}}AD 36\%\\ \color{red}{\underline{AD 28\%}}\end{tabular} & \begin{tabular}[c]{@{}l@{}}AD 25\%\\ \color{red}{\underline{AD 24\%}}\end{tabular} & \begin{tabular}[c]{@{}l@{}}AD \, 4\%\\ \color{red}{\underline{AD 23\%}}\end{tabular} & \begin{tabular}[c]{@{}l@{}}AD 13\%\\ \color{red}{\underline{AD 17\%}}\end{tabular} & \begin{tabular}[c]{@{}l@{}}AD 17\%\\ \color{red}{\underline{AD 20\%}}\end{tabular}  & \begin{tabular}[c]{@{}l@{}}AD 11\%\\ \color{red}{\underline{AD 23\%}}\end{tabular} \\
		&                                                                           &  &                                                           &                                                           &                                                            &                                                           &                                                            &                                                           \\
		&                                                                           &  & \begin{tabular}[c]{@{}l@{}}TA 28\%\\ \color{red}{\underline{TA 20\%}}\end{tabular} & \begin{tabular}[c]{@{}l@{}}TA 17\%\\ \color{red}{\underline{TA 19\%}}\end{tabular} & \begin{tabular}[c]{@{}l@{}}TA 13\%\\ \color{red}{\underline{TA 20\%}}\end{tabular}  & \begin{tabular}[c]{@{}l@{}}TA 26\%\\ \color{red}{\underline{TA 26\%}}\end{tabular} & \begin{tabular}[c]{@{}l@{}}TA 44\%\\ \color{red}{\underline{TA 49\%}}\end{tabular}  & \begin{tabular}[c]{@{}l@{}}TA 80\%\\ \color{red}{\underline{TA 41\%}}\end{tabular}
	\end{tabular}
	\caption{Inferred probability of results in the experiments reported by Hamburger \cite{hamburger1939development}. \color{red}{\underline{Red underlined}} \color{black} entries are inferred results, and black entries are reported results. Calculated by taking expectations over all configurations, with $\beta=1$.}
	\label{Ham2}
\end{table}

\begin{table}[]
	\begin{tabular}{lllllllll}
		&                                                                           &  &                                                           &                                                           & \multicolumn{2}{l}{Donor tissue}                                                                                       &                                                            &                                                           \\
		&                                                                           &  & LB1                                                       & LB2                                                       & LB3                                                        & LB4                                                       & LB5                                                        & LB6                                                       \\
		&                                                                           &  &                                                           &                                                           &                                                            &                                                           &                                                            &                                                           \\
		&                                                                           &  & \begin{tabular}[c]{@{}l@{}}ND 36\%\\ \color{red}{\underline{ND 57\%}}\end{tabular} & \begin{tabular}[c]{@{}l@{}}ND 58\%\\ \color{red}{\underline{ND 58\%}}\end{tabular} & \begin{tabular}[c]{@{}l@{}}ND 83\%\\ \color{red}{\underline{ND 59\%}}\end{tabular}  & \begin{tabular}[c]{@{}l@{}}ND 61\%\\ \color{red}{\underline{ND 60\%}}\end{tabular} & \begin{tabular}[c]{@{}l@{}}ND 39\% \\ \color{red}{\underline{ND 34\%}}\end{tabular} & \begin{tabular}[c]{@{}l@{}}ND \, 9\%\\ \color{red}{\underline{ND 39\%}}\end{tabular}  \\
		&                                                                           &  &                                                           &                                                           &                                                            &                                                           &                                                            &                                                           \\
		\begin{tabular}[c]{@{}l@{}}Host\\ tissue\end{tabular} & \begin{tabular}[c]{@{}l@{}}CB at the same\\ stage with donor\end{tabular} &  & \begin{tabular}[c]{@{}l@{}}AD 36\%\\ \color{red}{\underline{AD 26\%}}\end{tabular} & \begin{tabular}[c]{@{}l@{}}AD 25\%\\ \color{red}{\underline{AD 22\%}}\end{tabular} & \begin{tabular}[c]{@{}l@{}}AD \, 4\%\\ \color{red}{\underline{AD 20\%}}\end{tabular} & \begin{tabular}[c]{@{}l@{}}AD 13\%\\ \color{red}{\underline{AD 13\%}}\end{tabular} & \begin{tabular}[c]{@{}l@{}}AD 17\%\\ \color{red}{\underline{AD 15\%}}\end{tabular}  & \begin{tabular}[c]{@{}l@{}}AD 11\%\\ \color{red}{\underline{AD 18\%}}\end{tabular} \\
		&                                                                           &  &                                                           &                                                           &                                                            &                                                           &                                                            &                                                           \\
		&                                                                           &  & \begin{tabular}[c]{@{}l@{}}TA 28\%\\ \color{red}{\underline{TA 17\%}}\end{tabular} & \begin{tabular}[c]{@{}l@{}}TA 17\%\\ \color{red}{\underline{TA 20\%}}\end{tabular} & \begin{tabular}[c]{@{}l@{}}TA 13\%\\ \color{red}{\underline{TA 21\%}}\end{tabular}  & \begin{tabular}[c]{@{}l@{}}TA 26\%\\ \color{red}{\underline{TA 27\%}}\end{tabular} & \begin{tabular}[c]{@{}l@{}}TA 44\%\\ \color{red}{\underline{TA 51\%}}\end{tabular}  & \begin{tabular}[c]{@{}l@{}}TA 80\%\\ \color{red}{\underline{TA 43\%}}\end{tabular}
	\end{tabular}
	\caption{Inferred probability of results in the experiments reported by Hamburger \cite{hamburger1939development}. \color{red}{\underline{Red underlined}} \color{black} entries are inferred results, and black entries are reported results. Calculated by taking expectations over all configurations, with $\beta=2$.}
	\label{Ham3}
\end{table}

\section{Experimental design and inference}
\label{Snew}
Consider a group of tissue transplantation experiments, and we want to know all the experimental results. With our inference methods, we do not need to conduct all experiments, but only some of them, and use these conducted experiments to infer others. To have satisfactory inference results, each non-conducted experiment should have several (e.g., two) similar experiments that are conducted, since the inference uses the similarities between tissues. Assume we can choose which experiments to conduct, then the question is: how to choose experiments to conduct, so that the number of conducted experiments is minimized, meanwhile each non-conducted experiment is similar to at least $k$ conducted experiments. This is an experimental design problem.

Assume the experiment similarity chart is known. For each node (experiment) of this chart, we color it black (conducted) or white (non-conducted). Then the experimental design problem becomes a coloring problem: how to color a chart, so that each white node is neighboring to at least $k$ black nodes, and the number of black nodes is minimized?

For some cases, such as the experiments reported by Henry and Grainger \cite{henry1987inductive} (Table \ref{HG}), the experiment similarity chart is a subset of the two-dimensional square lattice $\mathbb{Z}^2$. In practice, each tissue transplantation experiment should be described by four components: donor tissue type, donor tissue developmental stage, host tissue type, host tissue developmental stage. If we slightly modify each component, we obtain a similar experiment. Thus the experiment similarity chart can be regarded as a subset of the four-dimensional square lattice $\mathbb{Z}^4$, where each experiment has a four-dimensional coordinate $(x_1,x_2,x_3,x_4)$. In general, if the experiment similarity chart is small and irregular, we can exhaustively search for the optimal design; if the experiment similarity chart is large, its structure should be close to a subset of $\mathbb{Z}^n$.

We directly consider the idealized general problem: Color each node of the $n$-dimensional square lattice $\mathbb{Z}^n$ black or white, so that each white node is neighboring to at least $k$ black node, and the number of black nodes is minimized. Here each node has a coordinate $(x_1,x_2,\ldots,x_n)$, and two nodes are neighboring if one component of their coordinates differs by $1$, and other components are equal.

Each white node is neighboring to at least $k$ black nodes, and each black node is neighboring to at most $2n$ white nodes. Therefore, the theoretical upper bound of white-black ratio is $2n:k$, and the minimal proportion of black nodes is $k/(2n+k)$. If this most efficient design in theory exists, then it should have the following properties: two black nodes are not neighboring, and each white node is neighboring to exactly $k$ black nodes. If $k$ is a divisor of $2n$, then such design exists. If $k$ is not a divisor of $2n$, then the white-black ratio is not an integer, and such design might not exist. We need to introduce a notation for congruence: $a\equiv b\text{ } (\text{mod }  c)$ means $c$ is a divisor of $a-b$. For example, $7\equiv 1\text{ } (\text{mod } 2)$.

\begin{proposition}
If $k$ is a divisor of $n$, color a node black if and only if its coordinate satisfies $a_1x_1+a_2x_2+\cdots+a_nx_n\equiv 0 \text{ } (\text{mod }  (2n/k)+1)$, where the coefficients $a_1,a_2,\ldots,a_n$ are $k$ groups of $1,2,\ldots,n/k$: $1,2,\ldots,n/k,1,2,\ldots,n/k,\ldots,1,2,\ldots,n/k$. Then black nodes are not neighboring, each white node is neighboring to $k$ black nodes, and the proportion of black nodes is $k/(2n+k)$.
\label{prop1}
\end{proposition}
\begin{proof}
A node is neighboring to $2n$ nodes. If the value of $a_1x_1+a_2x_2+\cdots+a_nx_n \text{ } (\text{mod }  (2n/k)+1)$ for this node is $i$, then the values of $a_1x_1+a_2x_2+\cdots+a_nx_n \text{ } (\text{mod }  (2n/k)+1)$ for its neighboring nodes are $k$ groups of $i+1,i+2,\ldots,i+n/k$ and $k$ groups of $i-1,i-2,\ldots,i-n/k$. If $i=0$, then none of these $2n$ values is $0$, meaning that black nodes are not neighboring; if $i\ne 0$, then exactly $k$ of these $2n$ values are $0$, meaning that each white node is neighboring to exactly $k$ black nodes.
\end{proof}

\begin{proposition}
If $k$ is a divisor of $2n$, but not a divisor of $n$, color a node black if and only if its coordinate satisfies $a_1x_1+a_2x_2+\cdots+a_nx_n\equiv 0 \text{ } (\text{mod }  (2n/k)+1)$, where the coefficients $a_1,a_2,\ldots,a_n$ are $k/2$ groups of $1,2,\ldots,2n/k$: $1,2,\ldots,2n/k,1,2,\ldots,2n/k,\ldots,1,2,\ldots,2n/k$. Then black nodes are not neighboring, each white node is neighboring to $k$ black nodes, and the proportion of black nodes is $k/(2n+k)$.
\label{prop2}
\end{proposition}
\begin{proof}
A node is neighboring to $2n$ nodes. If the value of $a_1x_1+a_2x_2+\cdots+a_nx_n \text{ } (\text{mod }  (2n/k)+1)$ for this node is $i$, then the values of $a_1x_1+a_2x_2+\cdots+a_nx_n \text{ } (\text{mod }  (2n/k)+1)$ for its neighboring nodes are $k/2$ groups of $i+1,i+2,\ldots,i+2n/k$ and $k/2$ groups of $i-1,i-2,\ldots,i-2n/k$, which are equivalent with $k$ groups of $i+1,i+2,\ldots,i+2n/k$. If $i=0$, then none of these $2n$ values is $0$, meaning that black nodes are not neighboring; if $i\ne 0$, then exactly $k$ of these $2n$ values are $0$, meaning that each white node is neighboring to exactly $k$ black nodes.
\end{proof}

When $n=2$, the black color condition for $k=4$ is $x_1+x_2\equiv 0 \text{ } (\text{mod } 2)$ ($1/2$ black nodes); the black color condition for $k=2$ is $x_1+x_2\equiv 0 \text{ } (\text{mod } 3)$ ($1/3$ black nodes); the black color condition for $k=1$ is $x_1+2x_2\equiv 0 \text{ } (\text{mod } 5)$ ($1/5$ black nodes). See Table \ref{co} for the visualized coloring methods.

When $n=4$, the black color condition for $k=8$ is $x_1+x_2+x_3+x_4\equiv 0 \text{ } (\text{mod } 2)$ ($1/2$ black nodes); the black color condition for $k=4$ is $x_1+x_2+x_3+x_4\equiv 0 \text{ } (\text{mod } 3)$ ($1/3$ black nodes); the black color condition for $k=2$ is $x_1+2x_2+x_3+2x_4\equiv 0 \text{ } (\text{mod } 5)$ ($1/5$ black nodes); the black color condition for $k=1$ is $x_1+2x_2+3x_3+4x_4\equiv 0 \text{ } (\text{mod } 9)$ ($1/9$ black nodes).

\begin{table}[]
\begin{tabular}{lllllllll}
     &                         &                        &                        &                        & \multicolumn{2}{l}{Donor}                       &                        &                        \\
     &                         & $T_1$                     & $T_2$                     & $T_3$                     & $T_4$                     & $T_5$                     & $T_6$                    & $T_7$                     \\ \cline{3-9} 
     & \multicolumn{1}{l|}{$T_1$} & \multicolumn{1}{l|}{\cellcolor{black}} & \multicolumn{1}{l|}{}  & \multicolumn{1}{l|}{\cellcolor{black}} & \multicolumn{1}{l|}{}  & \multicolumn{1}{l|}{\cellcolor{black}} & \multicolumn{1}{l|}{}  & \multicolumn{1}{l|}{\cellcolor{black}} \\ \cline{3-9} 
     & \multicolumn{1}{l|}{$T_2$} & \multicolumn{1}{l|}{}  & \multicolumn{1}{l|}{\cellcolor{black}} & \multicolumn{1}{l|}{}  & \multicolumn{1}{l|}{\cellcolor{black}} & \multicolumn{1}{l|}{}  & \multicolumn{1}{l|}{\cellcolor{black}} & \multicolumn{1}{l|}{}  \\ \cline{3-9} 
     & \multicolumn{1}{l|}{$T_3$} & \multicolumn{1}{l|}{\cellcolor{black}} & \multicolumn{1}{l|}{}  & \multicolumn{1}{l|}{\cellcolor{black}} & \multicolumn{1}{l|}{}  & \multicolumn{1}{l|}{\cellcolor{black}} & \multicolumn{1}{l|}{}  & \multicolumn{1}{l|}{\cellcolor{black}} \\ \cline{3-9} 
Host & \multicolumn{1}{l|}{$T_4$} & \multicolumn{1}{l|}{}  & \multicolumn{1}{l|}{\cellcolor{black}} & \multicolumn{1}{l|}{}  & \multicolumn{1}{l|}{\cellcolor{black}} & \multicolumn{1}{l|}{}  & \multicolumn{1}{l|}{\cellcolor{black}} & \multicolumn{1}{l|}{}  \\ \cline{3-9} 
     & \multicolumn{1}{l|}{$T_5$} & \multicolumn{1}{l|}{\cellcolor{black}} & \multicolumn{1}{l|}{}  & \multicolumn{1}{l|}{\cellcolor{black}} & \multicolumn{1}{l|}{}  & \multicolumn{1}{l|}{\cellcolor{black}} & \multicolumn{1}{l|}{}  & \multicolumn{1}{l|}{\cellcolor{black}} \\ \cline{3-9} 
     & \multicolumn{1}{l|}{$T_6$} & \multicolumn{1}{l|}{}  & \multicolumn{1}{l|}{\cellcolor{black}} & \multicolumn{1}{l|}{}  & \multicolumn{1}{l|}{\cellcolor{black}} & \multicolumn{1}{l|}{}  & \multicolumn{1}{l|}{\cellcolor{black}} & \multicolumn{1}{l|}{}  \\ \cline{3-9} 
     & \multicolumn{1}{l|}{$T_7$} & \multicolumn{1}{l|}{\cellcolor{black}} & \multicolumn{1}{l|}{}  & \multicolumn{1}{l|}{\cellcolor{black}} & \multicolumn{1}{l|}{}  & \multicolumn{1}{l|}{\cellcolor{black}} & \multicolumn{1}{l|}{}  & \multicolumn{1}{l|}{\cellcolor{black}} \\ \cline{3-9} 
     &                         &                        &                        &                        & \multicolumn{2}{l}{}                            &                        &                        \\
     &                         & T1                     & T2                     & T3                     & T4                     & T5                     & T6                     & T7                     \\ \cline{3-9} 
     & \multicolumn{1}{l|}{T1} & \multicolumn{1}{l|}{}  & \multicolumn{1}{l|}{\cellcolor{black}}& \multicolumn{1}{l|}{}  & \multicolumn{1}{l|}{}  & \multicolumn{1}{l|}{\cellcolor{black}}& \multicolumn{1}{l|}{}  & \multicolumn{1}{l|}{}  \\ \cline{3-9} 
     & \multicolumn{1}{l|}{T2} & \multicolumn{1}{l|}{\cellcolor{black}}& \multicolumn{1}{l|}{}  & \multicolumn{1}{l|}{}  & \multicolumn{1}{l|}{\cellcolor{black}}& \multicolumn{1}{l|}{}  & \multicolumn{1}{l|}{}  & \multicolumn{1}{l|}{\cellcolor{black}}\\ \cline{3-9} 
     & \multicolumn{1}{l|}{T3} & \multicolumn{1}{l|}{}  & \multicolumn{1}{l|}{}  & \multicolumn{1}{l|}{\cellcolor{black}}& \multicolumn{1}{l|}{}  & \multicolumn{1}{l|}{}  & \multicolumn{1}{l|}{\cellcolor{black}}& \multicolumn{1}{l|}{}  \\ \cline{3-9} 
Host & \multicolumn{1}{l|}{T4} & \multicolumn{1}{l|}{}  & \multicolumn{1}{l|}{\cellcolor{black}}& \multicolumn{1}{l|}{}  & \multicolumn{1}{l|}{}  & \multicolumn{1}{l|}{\cellcolor{black}}& \multicolumn{1}{l|}{}  & \multicolumn{1}{l|}{}  \\ \cline{3-9} 
     & \multicolumn{1}{l|}{T5} & \multicolumn{1}{l|}{\cellcolor{black}}& \multicolumn{1}{l|}{}  & \multicolumn{1}{l|}{}  & \multicolumn{1}{l|}{\cellcolor{black}}& \multicolumn{1}{l|}{}  & \multicolumn{1}{l|}{}  & \multicolumn{1}{l|}{\cellcolor{black}}\\ \cline{3-9} 
     & \multicolumn{1}{l|}{T6} & \multicolumn{1}{l|}{}  & \multicolumn{1}{l|}{}  & \multicolumn{1}{l|}{\cellcolor{black}}& \multicolumn{1}{l|}{}  & \multicolumn{1}{l|}{}  & \multicolumn{1}{l|}{\cellcolor{black}}& \multicolumn{1}{l|}{}  \\ \cline{3-9} 
     & \multicolumn{1}{l|}{T7} & \multicolumn{1}{l|}{}  & \multicolumn{1}{l|}{\cellcolor{black}}& \multicolumn{1}{l|}{}  & \multicolumn{1}{l|}{}  & \multicolumn{1}{l|}{\cellcolor{black}}& \multicolumn{1}{l|}{}  & \multicolumn{1}{l|}{}  \\ \cline{3-9} 
     &                         &                        &                        &                        & \multicolumn{2}{l}{}                            &                        &                        \\
     &                         & T1                     & T2                     & T3                     & T4                     & T5                     & T6                     & T7                     \\ \cline{3-9} 
     & \multicolumn{1}{l|}{T1} & \multicolumn{1}{l|}{}  & \multicolumn{1}{l|}{}  & \multicolumn{1}{l|}{\cellcolor{black}}& \multicolumn{1}{l|}{}  & \multicolumn{1}{l|}{}  & \multicolumn{1}{l|}{}  & \multicolumn{1}{l|}{}  \\ \cline{3-9} 
     & \multicolumn{1}{l|}{T2} & \multicolumn{1}{l|}{\cellcolor{black}}& \multicolumn{1}{l|}{}  & \multicolumn{1}{l|}{}  & \multicolumn{1}{l|}{}  & \multicolumn{1}{l|}{}  & \multicolumn{1}{l|}{\cellcolor{black}}& \multicolumn{1}{l|}{}  \\ \cline{3-9} 
     & \multicolumn{1}{l|}{T3} & \multicolumn{1}{l|}{}  & \multicolumn{1}{l|}{}  & \multicolumn{1}{l|}{}  & \multicolumn{1}{l|}{\cellcolor{black}}& \multicolumn{1}{l|}{}  & \multicolumn{1}{l|}{}  & \multicolumn{1}{l|}{}  \\ \cline{3-9} 
Host & \multicolumn{1}{l|}{T4} & \multicolumn{1}{l|}{}  & \multicolumn{1}{l|}{\cellcolor{black}}& \multicolumn{1}{l|}{}  & \multicolumn{1}{l|}{}  & \multicolumn{1}{l|}{}  & \multicolumn{1}{l|}{}  & \multicolumn{1}{l|}{\cellcolor{black}}\\ \cline{3-9} 
     & \multicolumn{1}{l|}{T5} & \multicolumn{1}{l|}{}  & \multicolumn{1}{l|}{}  & \multicolumn{1}{l|}{}  & \multicolumn{1}{l|}{}  & \multicolumn{1}{l|}{\cellcolor{black}}& \multicolumn{1}{l|}{}  & \multicolumn{1}{l|}{}  \\ \cline{3-9} 
     & \multicolumn{1}{l|}{T6} & \multicolumn{1}{l|}{}  & \multicolumn{1}{l|}{}  & \multicolumn{1}{l|}{\cellcolor{black}}& \multicolumn{1}{l|}{}  & \multicolumn{1}{l|}{}  & \multicolumn{1}{l|}{}  & \multicolumn{1}{l|}{}  \\ \cline{3-9} 
     & \multicolumn{1}{l|}{T7} & \multicolumn{1}{l|}{\cellcolor{black}}& \multicolumn{1}{l|}{}  & \multicolumn{1}{l|}{}  & \multicolumn{1}{l|}{}  & \multicolumn{1}{l|}{}  & \multicolumn{1}{l|}{\cellcolor{black}}& \multicolumn{1}{l|}{}  \\ \cline{3-9} 
\end{tabular}
\caption{Experimental designs in different cases, corresponding to two-dimensional experiment similarity chart $\mathbb{Z}^2$. Each cell is an experiment, and neighboring cells are similar experiments. Black cells are conducted experiments, and white cells are non-conducted experiments. Black cells are not neighboring. Each non-boundary white cell is neighboring to $k$ black cells, where $k=4$ (upper), $k=2$ (middle), $k=1$ (lower).}
\label{co}
\end{table}

We explicitly verify the construction for $n=4$, $k=1$. For a node with coordinate $(x_1,x_2,x_3,x_4)$, if $x_1+2x_2+3x_3+4x_4\equiv 0 \text{ } (\text{mod } 9)$, then none of its neighboring nodes is black. If $x_1+2x_2+3x_3+4x_4\equiv 1 \text{ } (\text{mod } 9)$, then $(x_1-1,x_2,x_3,x_4)$ is black, since $(x_1-1)+2x_2+3x_3+4x_4\equiv 0 \text{ } (\text{mod } 9)$. If $x_1+2x_2+3x_3+4x_4\equiv 2 \text{ } (\text{mod } 9)$, then $(x_1,x_2-1,x_3,x_4)$ is black, since $x_1+2(x_2-1)+3x_3+4x_4\equiv 0 \text{ } (\text{mod } 9)$. If $x_1+2x_2+3x_3+4x_4\equiv 3 \text{ } (\text{mod } 9)$, then $(x_1,x_2,x_3-1,x_4)$ is black, since $x_1+2x_2+3(x_3-1)+4x_4\equiv 0 \text{ } (\text{mod } 9)$. If $x_1+2x_2+3x_3+4x_4\equiv 4 \text{ } (\text{mod } 9)$, then $(x_1,x_2,x_3,x_4-1)$ is black, since $x_1+2x_2+3x_3+4(x_4-1)\equiv 0 \text{ } (\text{mod } 9)$. 
If $x_1+2x_2+3x_3+4x_4\equiv 5 \text{ } (\text{mod } 9)$, then $(x_1,x_2,x_3,x_4+1)$ is black, since $x_1+2x_2+3x_3+4(x_4+1)\equiv 0 \text{ } (\text{mod } 9)$. If $x_1+2x_2+3x_3+4x_4\equiv 6 \text{ } (\text{mod } 9)$, then $(x_1,x_2,x_3+1,x_4)$ is black, since $x_1+2x_2+3(x_3+1)+4x_4\equiv 0 \text{ } (\text{mod } 9)$. If $x_1+2x_2+3x_3+4x_4\equiv 7 \text{ } (\text{mod } 9)$, then $(x_1,x_2+1,x_3,x_4)$ is black, since $x_1+2(x_2+1)+3x_3+4x_4\equiv 0 \text{ } (\text{mod } 9)$. If $x_1+2x_2+3x_3+4x_4\equiv 8 \text{ } (\text{mod } 9)$, then $(x_1+1,x_2,x_3,x_4)$ is black, since $(x_1+1)+2x_2+3x_3+4x_4\equiv 0 \text{ } (\text{mod } 9)$. For this case, one white node corresponds to one black node, and one black node corresponds to eight white nodes, making the white-black ratio 8:1. That is why we use (mod $9$).

Our experience shows that $k=2$ should be enough to conduct plausible inferences. Therefore, when the experiment similarity chart is two- or four-dimensional, we only need to conduct $1/3$ or $1/5$ experiments to infer other experiments. The minimal requirement on the proportion of conducted experiments, corresponding to $k=1$, is $1/5$ ($\mathbb{Z}^2$) or $1/9$ ($\mathbb{Z}^4$). Smaller proportions might produce unreliable results. When the experiment similarity chart is more complicated, meaning that there are more similarity relationships, the number of experiments we need to conduct is even smaller, since the same known result can provide more information about its unknown neighbors.

\section{Discussion}
\label{S6}
In this paper, we summarize tissue transplantation experiments for various species and develop methods to infer the unknown experimental results in different cases. For each case, we conduct our inference methods with different values of parameters to show that we do not need fine-tuning with parameters (or similarity charts) to produce reasonable inference results.

We only apply our methods to the first three experiments in Section \ref{S2}. For the experiments reported by Jones and Woodland \cite{jones1987development} (Table \ref{Jo}), there are many experiments with unknown results that are not similar to any experiments with known results; therefore our methods fail to produce reliable inference results. For the experiments reported by Arresta et al. \cite{arresta2005lens} (Table \ref{AB}), Elliott et al. \cite{elliott2013transplantation} (Table \ref{EH}), and Kao and Chang \cite{kao1996homeotic} (Table \ref{Kao}), donor tissues are not similar, therefore experiments are not similar either, and our methods cannot be performed.

Our methods rely on the similarities between experiments. Therefore, to infer the result of one experiment, we need to know the results of some similar experiments. For example, the \emph{Xenopus laevis} experiments reported by Krneta-Stankic et al. \cite{krneta2010temporal} consider lip and mesoderm tissues, while the \emph{Xenopus laevis} experiments reported by Henry and Grainger \cite{henry1987inductive} consider ectoderm tissues. We do not have any information about the transplantation between lip/mesoderm and ectoderm tissues; thus, our methods fail to provide any inference on such experiments. This is why we cannot unify all the available experiments of the same species into a single table. Besides, how to determine such similarities between tissues/experiments, in the sense of conducting inference, is an essential problem. This requires a more fundamental understanding of various species.

We need many more experiments to verify the proper values of parameters $J_0,h_0,\beta$ in our method. These parameters are currently taken somewhat arbitrarily; thus, the results of our methods are not accurate, except the case in Section \ref{S5}. Besides, there are other choices of penalty functions and probability functions.

Once the factors of our methods have been determined, we could learn tissue transplantation with a relatively low cost: conducting a small portion of experiments is enough to infer all the experiments. With enough cumulated data, one can even develop a method to infer experimental results directly from features of donor and host tissues. Then measuring the properties of $n$ tissues is enough to infer $n\times n$ experiments. With that, we can enter a much higher level of understanding of tissue transplantation.

Our methods should not be limited to tissue transplantation experiments. There could be other biological problems that fit our method. A problem that fits our method needs to have nominal entries, some known and some unknown, and their similarities are essential. Besides, prior knowledge should be relatively limited, otherwise there could be more advanced tools.

\section*{Acknowledgements}
We would like to thank Dr. Simon Fourquet, Dr. Jie Ren, Dr. Alen Tosenberger, and Dr. Zikun Wang for helpful suggestions, and Prof. Mikhail Gromov, Prof. Robert Penner, and Prof. Hong Qian for fruitful discussions.

The work of Yue Wang and J\'er\'emie Kropp was supported by the Simons Foundation (IH\'ES program of mathematical biology). The work of Nadya Morozova was carried out within the framework of the state assignment to Komarov Botanical Institute RAS No. AAAA-A18-118051590122-8.

\section*{Declaration of interest}
The authors declare no conflict of interest.

\bibliographystyle{plain}
\bibliography{Tissue_Transplantation}
\end{document}